\documentclass[paper=A4,pagesize,ngerman,UKenglish,13pt]{scrartcl}
\usepackage{hyperref}
\usepackage[utf8]{inputenc}
\usepackage[ngerman,UKenglish]{babel}
\usepackage{csquotes}
\usepackage{graphicx}
\usepackage{amsmath}
\usepackage{amssymb}
\usepackage[tikz]{bclogo}
\usepackage{etoolbox}%
\usepackage{blindtext}%
\usepackage[skins,breakable,xparse]{tcolorbox}%
\usepackage{url}
\usepackage{lmodern}
\usepackage[T1]{fontenc}
\usepackage{microtype}
\usepackage{xcolor}
\usepackage{calc}
\usepackage{dsfont}
\KOMAoptions{DIV=14}%
\usepackage{subcaption}

\renewcommand{\tt}{\normalfont \ttfamily}

\newlength{\unit}
\usepackage{pdflscape}

\definecolor{bggrey}{RGB}{245,245,245}
\definecolor{ttblue}{RGB}{91,194,224}

\DeclareTotalTColorBox[auto counter]{\insertTP2}{ O{} m }
{ enhanced,
  boxrule=0pt,boxsep=0pt,arc=2mm,toptitle=3mm,top=3mm,left=7mm,right=1mm,pad at break=2mm,
  colframe=orange!10!white,interior hidden,
  coltitle=black,fonttitle=\bfseries\large,title={#1},
  overlay unbroken and first={\node[inner sep=0pt] (logo) at ([xshift=4mm,yshift=-5mm]frame.north west) {
};
    \draw[red,line width=3.5pt] (logo) -- ([xshift=4mm,yshift=1.5mm]frame.south west);  },
  overlay middle and last={\draw[red,line width=3.5pt] ([xshift=4mm,yshift=-1.5mm]frame.north west) -- ([xshift=4mm,yshift=1.5mm]frame.south west); },
  #1}
  {#2
  }

\tcbset{colback=black!5!white,colframe=black!75!black}
\newcommand{\insertTP}[2]{
  \begin{tcolorbox}[leftrule=0mm,rightrule=0mm,toprule=0mm,bottomrule=0mm]
    \hspace*{2pt}\parbox{4ex}{
}\hspace*{-5pt}
    \parbox{\textwidth-5ex}{
      \sffamily\bfseries
      #1}\\[0.5mm]
    #2
  \end{tcolorbox}
}

\usepackage[%
  automark,
  headsepline,                
]{scrpage2}

\lohead{\normalfont Section~\headmark}    
\rohead{\pagemark}                        
\automark[subsection]{section}
\setkomafont{title}{\sffamily\bfseries}
\setkomafont{caption}{\footnotesize\slshape}
\setkomafont{captionlabel}{\normalfont\bfseries}
\setcapindent{0pt}

\let\olddefparagraph\paragraph
\renewcommand{\paragraph}[1]{\olddefparagraph{#1}\mbox{}\\}

\newcommand{\col}{~,}
\newcommand{\pnt}{~.}
\newcommand{\T}{{\operatorname{t}}}

\DeclareMathOperator{\sgn}{sgn}

\definecolor{refcolor}{cmyk}{0.5,0,1,0}
\definecolor{darkred}{cmyk}{1,1,1,0.5}
\definecolor{darkviolet}{cmyk}{0.5,1,0,0.25}
\definecolor{lightviolet}{cmyk}{0.1,0.2,0,0.05}

\definecolor{mygreen}{rgb}{0,0.6,0}
\definecolor{mygray}{rgb}{0.5,0.5,0.5}
\definecolor{mymauve}{rgb}{0.58,0,0.82}
\definecolor{gray}{rgb}{0.4,0.4,0.4}
\definecolor{darkblue}{rgb}{0.0,0.0,0.6}
\definecolor{cyan}{rgb}{0.0,0.6,0.6}
\definecolor{lila}{rgb}{0.45,0.45,0.8}
\definecolor{pink}{rgb}{1.0,0,1.0}
\definecolor{background}{rgb}{0.95,0.95,0.95}

\definecolor{comment}{rgb}{0.2,0.7,0.7}
\definecolor{keyword}{rgb}{0.5,0,0.48}

\begin{document}

\title{%
  \begin{flushright}\vspace*{-2cm}
  \end{flushright}\vspace*{-0.5cm}
  AWEsome:\\ An open-source test platform for airborne wind energy systems \\[2mm]}
\author{
\large
Philip Bechtle$^a$, 
Thomas Gehrmann$^b$, 
Christoph Sieg$^c$, 
Udo Zillmann$^d$ \footnote{Author names are in alphabetical order. \newline
\hspace{6.3mm} $^a$ Universität Bonn, \texttt{bechtle@physik.uni-bonn.de} \newline
\hspace{6.3mm} $^b$ Universität Bonn, \texttt{thomasgehrmann@gmx.net} \newline
\hspace{6.3mm} $^c$ Humboldt-Universität zu Berlin, \texttt{csieg@physik.hu-berlin.de} \newline
\hspace{6.3mm} $^d$ Daidalos Capital, \texttt{zillmann@daidalos-capital.com}\newline
\hspace{6.3mm} \url{awesome.physik.uni-bonn.de} \newline
}
}

\date{\large\today{}}
\maketitle
\setcounter{tocdepth}{2}

\begin{center}
Abstract
\end{center}
In this paper we present AWEsome (Airborne Wind Energy Standardized
Open-source Model Environment), a test platform for airborne wind energy systems
that consists of low-cost hardware and is entirely based on open-source 
software. It can hence be used without the need of large financial
investments, in particular by research groups and startups to acquire
first experiences in their flight operations, to test novel control
strategies or technical designs, or for usage in public relations.
Our system consists of a modified off-the-shelf model aircraft that is
controlled by the {\tt pixhawk} autopilot hardware and the {\tt ardupilot}
software for fixed wing aircraft. The aircraft is attached to the ground 
by a tether. We have implemented new flight modes 
for the autonomous tethered flight of the aircraft along periodic
patterns. We present the principal functionality of our algorithms. 
We report on first successful tests of these modes in real flights.

\clearpage
\pagestyle{scrheadings}

\section{Introduction}\label{sec:Intro}

Airborne wind energy (AWE) systems are devices that convert wind energy 
into mechanical and ultimately electrical power via an aerodynamically 
active part that is at most flexibly connected to a ground station
via one or several tethers. The lift force acting on the airborne part 
for compensating the gravitational force can either be a static 
buoyant force generated by lighter-than-air structures such as balloons 
or it can be dynamically generated in cross-wind flight. 

The concept of extracting wind energy via an airborne system 
in cross-wind flight has been 
investigated more than 35 years ago by Loyd in his
seminal paper \cite{LOYD1980}. 
Loyd determined the amount of mechanical power an aircraft 
that is attached to the ground via a tether
can in principle 
harvest from a wind field.
The maximal mechanical power output 
is given by
\begin{equation}\label{Pmax}
  P_{\text{max}}\lesssim \frac{2}{27}\rho A
  v_{\text{W}}^3C_{\text{L}}\left(\frac{C_{\text{L}}}{C_{\text{D}}}\right)^2
\cos^3\gamma
\pnt
\end{equation}
Here, $A$, $C_{\text{L}}$ and $C_{\text{D}}$ are the wing area, lift
and drag coefficients of the aircraft, respectively. Moreover,
$v_{\text{w}}$ is the wind velocity and $\rho$ is the air density. In
addition, the effect of a non-vanishing inclination angle $\gamma$
that is the angle between the wind velocity vector and the vector that
points along a straight tether from the attachment point on the ground
to the aircraft has been considered. 
In case of 
a horizontal wind field, this angle is identical to the elevation angle
of the ideal (straight) tether connecting the ground station and aircraft.
The maximal
power \eqref{Pmax} is reached for a particular value for the 
apparent air speed $v_{\text{a}}$ felt by the aircraft. This can be achieved
by adjusting the drag coefficients of 
the on-board generators in case of drag-mode operation
and the reel-out speed of the tether in case of a pumping mode operation, 
respectively. 

For aircraft with rigid wings which have $C_{\text{L}}\simeq 1$ and
$\frac{C_{\text{L}}}{C_{\text{D}}}\gg1$ the possible power output is
surprisingly high \cite{Ahrens2013}.
These and further advantages such as the possibility to harvest
steadier and stronger winds at higher altitudes and the reduction of
material costs compared to conventional wind turbines make AWE a very
promising candidate for contributing significantly to solving the
world's sustainable energy problem~\cite{Ahrens2013}. Vivid research
and development of AWE systems is already performed by academic
research groups and startup companies; see e.g.\ \cite{Ahrens2013} for
an overview of activities. A comprehensive review of the different
design approaches is also given in~\cite{Cherubini20151461,EbrahimiSalari2017}.
A short introduction into the economic viability is given in \cite{Zillmann2016}.

Especially the need to design and build the required hardware and to
potentially damage or even lose it in field tests impose high economic
risks on small research groups and companies.  Besides that, there
exist various different strategies and guidelines especially for the
aircraft design and power conversion, and it is not yet clear which of
them maximizes the efficiency of the system.

The purpose of our research is to provide an open-source (OS)
cost-efficient and hence, in case of a failure, disposable test
platform for AWE systems. We call it AWEsome, the Airborne Wind Energy
Standardized Open-source Model Environment. This platform enables especially
teams with small financial resources to start gaining experience in
autonomous flight operations, test their deployment strategy, realize
various designs in hardware or software, and to test them even
allowing the risk of total loss.\footnote{Similar in spirit, another
  test platform based on a model plane is presented in
  \cite{DBLP:journals/corr/FagianoVRSO16} for testing autonomous
  take-off.}

In addition to the abovementioned consideration, the OS approach
offers considerable potential advantages for the AWE industy as a
whole. The development of safe, certified and efficient control
algorithms is amongst the key prerequisites 
to finally commercialize AWE systems.
The scrutiny offered by the OS approach should be seen as an important
asset for maximizing safety, reliability and efficiency. We also believe
that as long as the most efficient design has not 
been identified, close collaboration and exchange of results
will accelerate the technological development and hence be 
advantageous for all companies for developing their individual commercial
systems.

One of the key aspects of this paper is the description of the
OS autopilot which allows autonomous flight of the AWEsome system, 
and of the theoretical foundations and embedding of the control algorithms
which are used to navigate along the desired periodic flight path. 
Besides the basic
control strategy which is suitable for the limited
computing capacity of the \texttt{pixhawk} microcontroller and 
which is described in this paper, a lot of research has been invested into
the control of different types of cross-wind AWE systems. This research can be
classified in several ways, for instance according to the location 
of the control system and actuators.
Some approaches have ground-based controllers and actuators that control the airborne part via multiple tethers (see e.g.~\cite{Bormann2013}) or in addition an airborne actuator (see e.g.~\cite{Luchsinger2015}). They require models 
for the tethers. Others have a separate airborne control unit from which 
a bridle originates for the control of a soft wing (see e.g.\ \cite{Fechner2012}). Furthermore, the control unit and actuators can be part of a (rigid)
aircraft to which only a single tether is attached 
(see e.g.\ \cite{Ruiterkamp2013}). 
Also, we can classify different approaches according to their employed control 
strategies. For instance, optimal control has been applied for maximizing the average power output \cite{Houska2007}, in addition also considering the electrical power conversion \cite{Stuyts2015}. Approaches based on nonlinear model predictive control are presented in \cite{zanon:hal-00916755,WooEtal:2017:IFA_5622}.
For recent developments concerning further more advanced control strategies
of AWE systems, see 
also \cite{Terink2011,2012arXiv1202.3641E,Erhard2017,Joerger2016,Costello2017,DBLP:journals/corr/abs-1301-1064,Nguyen:2015,7525563,1742-6596-524-1-012081}. However, in this paper we will stick to a simple proportional-differential (PD) controller for lateral navigation. 
Our aim is to demonstrate that even an elementary setup is sufficient for
tethered flight and hence is a suitable intuitively understandable starting 
point for developing and testing various aspects of AWE systems. 

In this paper, we will present our design platform, introduce our
concepts on which our implementations of flight modes at a tether of
constant length rely and discuss first results from field tests.
Finally, we will give an outlook for the further development of the
platform.  Some complementary details can also be found in the master
thesis of one of us \cite{Gehrmann:2016}.
 
\section{Foundations of the test platform}

In the following, we describe the hard- and software of the test platform 
and its modifications. Moreover, we summarize the foundations on which 
our algorithms for tethered flight modes are based.  
Some complementary details especially concerning the hardware modifications
and the implementation of the algorithms can be
found in the master thesis of one of us \cite{Gehrmann:2016}. 
Throughout the text we use the {\tt typewriter} font to indicate proper names 
of hard- and software and of functions and variables of the source code.

\subsection{Hard- and Software}

\begin{figure}[ht]
    \centering
    \def\svgwidth{\textwidth}
    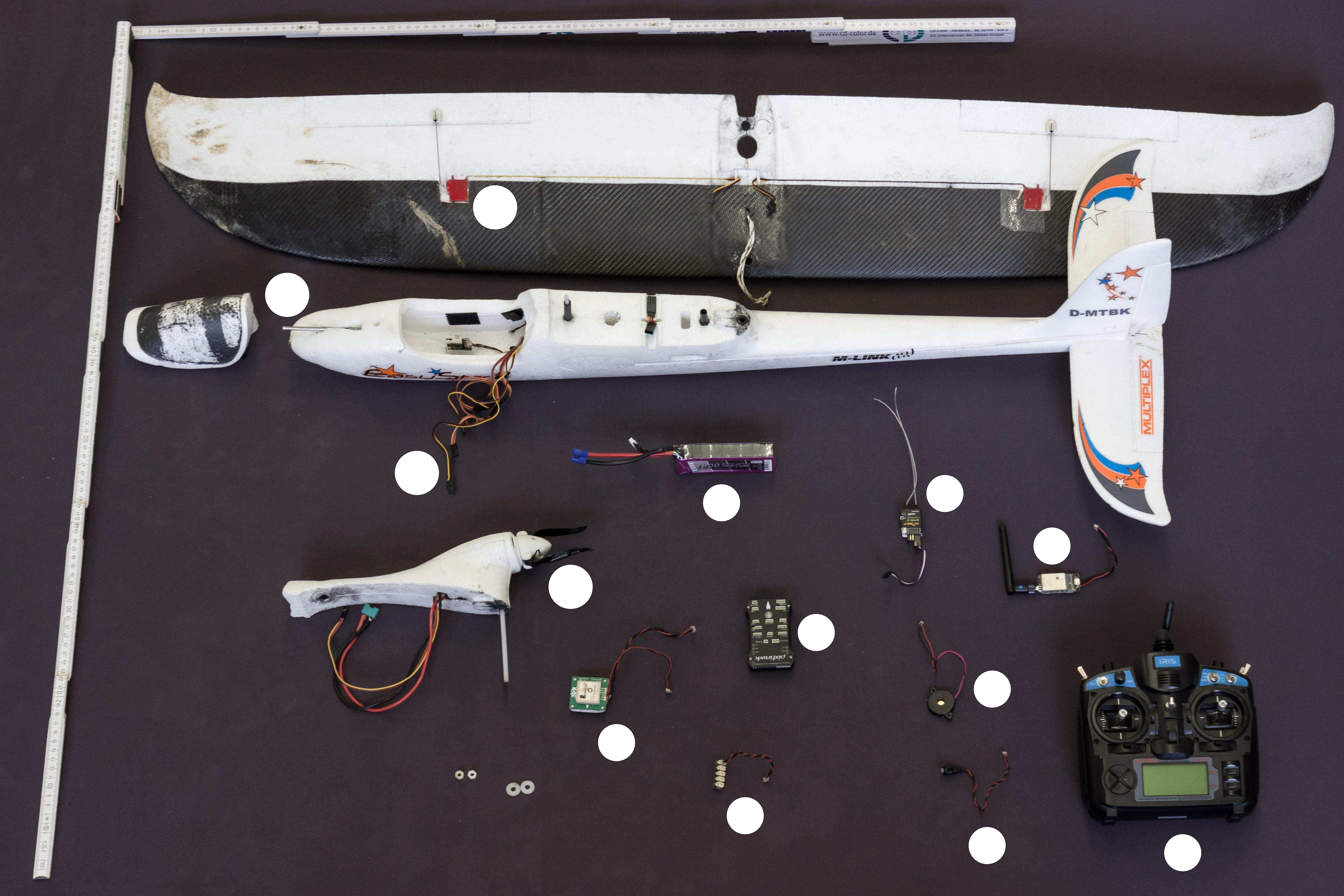
\caption{Components of the modified aircraft. \\
\textbf{1 transmitter} for remote radio control,  
\textbf{2 receiver} to receive signals from the transmitter, \\
\textbf{3 telemetry} used for communication between the GCS and aircraft, \\
\textbf{4 buzzer} makes sounds to inform about status, 
\textbf{5 safety switch} to prevent from accidental arming, \\
    \textbf{6 I$^2$C splitter} provides three additional ports for digital peripherals, \\
\textbf{7 GPS/compass module} provides positioning and heading data, \\
\textbf{8 pixhawk} microcontroller including acceleration sensors, gyroscopes, magnetometers, barometers\\
\textbf{9 battery} provides power, 
\textbf{10 propeller} provides thrust, 
\textbf{11 connection cables} to servos, \\
\textbf{12 airspeed sensor} measures apparent airspeed, \\
\textbf{13 servos} to steer the ailerons; servos for rudder and elevator are inside the fuselage \\
\label{fig:components}
}
\end{figure}

The airborne part of the test platform is an off-the-shelf polystyrene model aircraft, the Easy Star II\footnote{See e.g.\ {\tt https://www.multiplex-rc.de/produkte/264260-rr-easystar-ii-mit-bl-antrieb}}, which we modified for tethered flight. Its control surfaces are the flaps, ailerons and rudder, and it has an electric throttle. We reinforced the aircraft with carbon fabric of specific weight $160\text{g}/\text{m}^2$ that is laminated onto
the wings and bottom of the fuselage with epoxy resin. Thereby, the separate left and right wing have been glued together to a single indecomposable structure.
This ensures that the aircraft can sustain the tethered flight in which the tether force acts like a payload excess. The tether is tied to a 
carbon tube that sits inside the wings and replaces the fiberglass tube
originally delivered with the model. Moreover, the tube rests on carbon ribs that we have placed into the wings in order to transfer the force 
to the carbon fabric. At the expense of only adding about 150 grams 
of extra mass, the carbon structures significantly increase the 
stiffness of the aircraft. Moreover, the forces that the reinforced 
aircraft can sustain can be at least an order of magnitude higher than
those admissible for the aircraft without modifications. We checked this 
with several material samples. More details of the modifications can be 
found in the master thesis of one of us \cite{Gehrmann:2016}. 
A picture of the components is presented in figure \ref{fig:components}.

The ground station is currently a simple fishing rod with an offshore 
fising reel mounted, which serves 
for manually reeling in and out the tether. 
In comparison to a standard fishing reel, the offshore fishing reel 
has the advantage that its reeling technique does not twist the tether.

The autopilot that controls the aircraft 
consists of the {\tt pixhawk} circuit board \cite{meier2015px4}
and the {\tt ardupilot} software \cite{ardupilot}, release 
3.6.0 of June 6, 2016. The {\tt pixhawk} 
contains accelerometers, gyroscopes, magnetometers and a barometer.  
A GPS receiver and an airspeed sensor are externally connected and
communicate with the {\tt pixhawk} via the I$^2$C bus \cite{I2C}. For manual flight, an RC receiver is also connected to the {\tt pixhawk}.
Moreover, a WLAN transmitter employs the micro air vehicle communication protocol {\tt MAVLink} \cite{MAVLink} to send and receive telemetry data
to and respectively from a ground control station (GCS).

The {\tt ardupilot} software is written in {\tt C++}, and it 
encompasses vehicle specific codes 
for copters, rovers and fixed-wing aircraft. The code for fixed-wing aircraft 
that is the relevant one here is denoted as {\tt ArduPlane}. Each 
vehicle-specific code can be compiled for
different target hardware, including the {\tt pixhawk}, and also 
for a software in the loop (SITL) 
target. The latter allows to test the software e.g.\ in the OS flight dynamics
model {\tt JSBSim} \cite{JSBSim}. 

Finally, the GCS is a laptop with {\tt Mission Planner} \cite{missionplanner}
or {\tt APM Planner} \cite{APMPlanner} running on its {\tt Windows} or {\tt Linux} operating system, respectively. For SITL simulations, {\tt JSBSim} uses {\tt MAVProxy} \cite{MAVProxy} as GCS.

\subsection{Foundations of the tethered flight patterns}

Suppose the aircraft is attached via a tether to a fixed
point on the ground, denoted as `home' location in the following. The tether of cross section $A_{\text{t}}$ shall have a fixed length $l_0$ if no force is applied, and its length $l$ and stress $\tau$
shall increase linearly with the applied tether force $F_{\text{t}}$. The tether force and stress as a function of the length then read
\begin{equation}\label{tetherforce}
F_{\text{t}}=K_{\text{t}}(l-l_0)\theta(l-l_0)\col\qquad 
\tau=\frac{F_{\text{t}}}{A_{\text{t}}}
\col
\end{equation}
where $K_{\text{t}}$ is a constant depending on the material and $\theta$ is the Heavyside step function.

An aircraft that flies attached to such a tether and keeps 
$\tau>0$ constant is confined to a hemisphere 
of radius $R=l(\tau)=l_0+\frac{A_{\text{t}}}{K_{\text{t}}}\tau$
that is centered around the `home' location. Hence, all flight paths that 
are subject to this condition are curves on that hemisphere. What is important
for the later analysis is that the simple model \eqref{tetherforce} incorporates
the fact that at constant tether length  
the distance of the aircraft's position should be a measure for the tether stress and hence the possible power output. This allows us to make some
qualitative statements about the tether force even in the present case where
a direct measurement of that force is not yet implemented.
Of course, the simple relation \eqref{tetherforce} that is based on Hook's law for ideal springs, is only an idealization. 
A more realistic model should also consider that the drag force that acts 
when the tether is moved through the air and 
the gravitational force let the tether run along a curve rather than a straight line, thereby decreasing the distance of the aircraft to the `home'
location at constant tether length and stress. Moreover, 
accelerations of the aircraft may excite oscillations of the tether. 
Here, we will refrain from including these effects.

\begin{figure}[t]
\centering
\includegraphics[width=0.4\textwidth,trim = 50mm 30mm 35mm 75mm, clip]{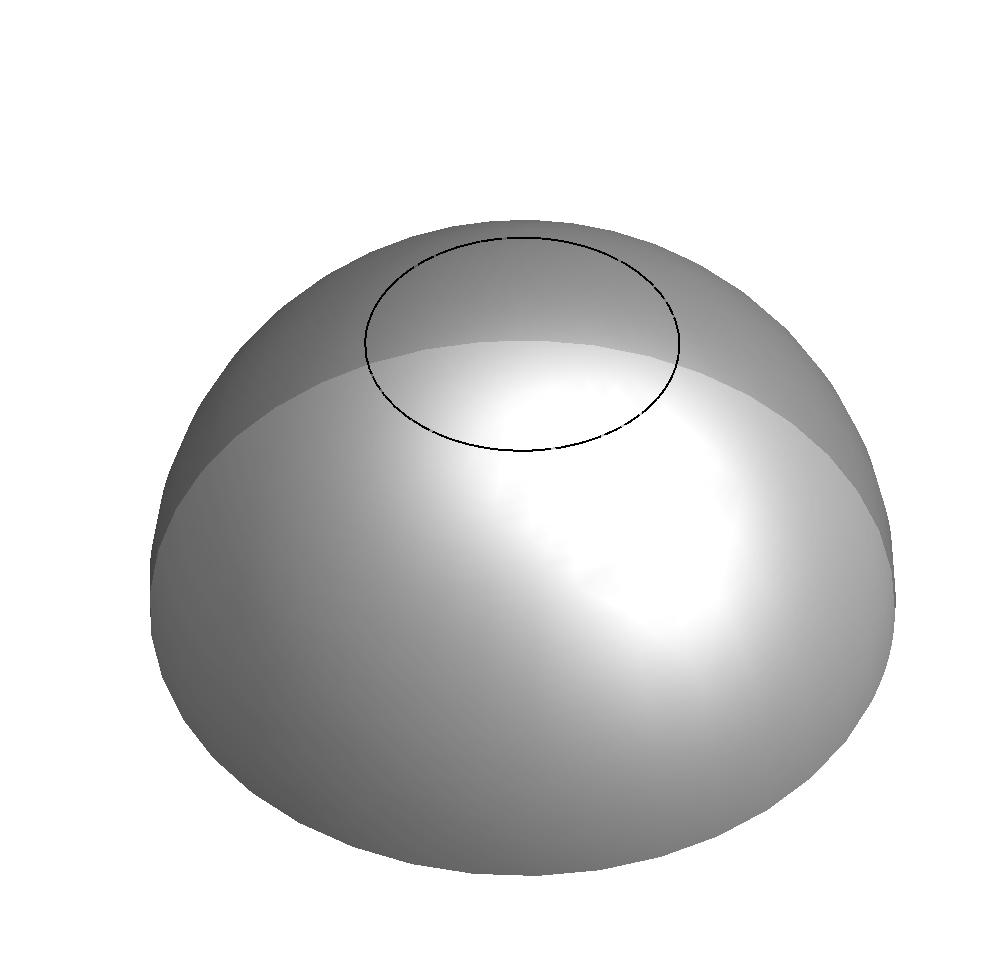}\qquad%
\includegraphics[width=0.4\textwidth,trim = 50mm 30mm 35mm 75mm, clip]{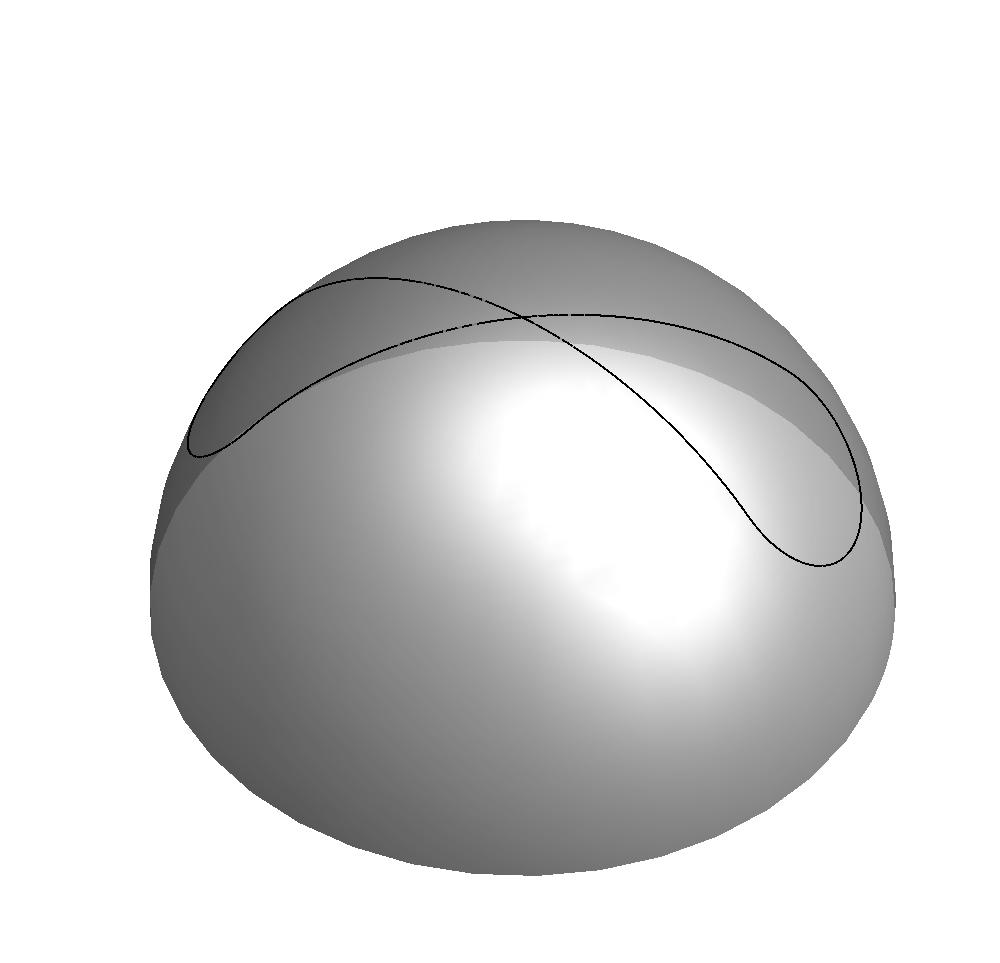}
\caption{Small circle and figure-eight pattern as simple periodic flight paths on a hemisphere for elevation $\gamma=45^\circ$.
\label{fig:flightcircleeight}}
\end{figure}

The simplest periodic curve on a hemisphere is a circle.
However, flying along a circle leads to a cumulation of twist on the tether.
In order to avoid this, the aircraft can e.g.\ fly along a curve consisting 
of two (circular) segments of opposite orientation,
resembling e.g the figure eight. 
Both these periodic curves, the circle and figure-eight pattern on the hemisphere, are shown in figure
\ref{fig:flightcircleeight}.
The circle and figure-eight pattern have thereby been placed horizontally, i.e.\ with an 
elevation angle $\gamma=\frac{\pi}{2}$ of the direction vector pointing 
from the center of the sphere to the center of the circle or 
crossing point of the figure-eight pattern. 

A generic (inclined) and oriented version of the small circle of 
figure \ref{fig:flightcircleeight} on a hemisphere of radius $R$ centered around
the origin can be 
uniquely specified by (half of) the 
opening angle $0\le \theta_\rho\le\frac{\pi}{2}$, the
elevation and azimuthal angle $0\le\gamma_{\text{c}}\le\frac{\pi}{2}$ and $0\le\psi_{\text{c}}\le2\pi$, respectively, and the orientation $\sigma=\pm 1$.
The angles define the circle radius $0\le R_{\text{c}}\le R$ and a unit vector as
\begin{equation}\label{Rcercdef}
R_{\text{c}}=R\sin\theta_\rho
\col\qquad
\vec e_{\text{r}_{\text{c}}}
=\begin{pmatrix}
\sin\theta_{\text{c}}\cos\psi_{\text{c}} \\
\sin\theta_{\text{c}}\sin\psi_{\text{c}} \\
-\cos\theta_{\text{c}}
\end{pmatrix}
\col\qquad
\theta_{\text{c}}=\frac{\pi}{2}-\gamma_{\text{c}}
\pnt
\end{equation}
The vector is given in the north-east-down (NED) coordinate system of avionics,
where vectors with positive third components point downwards.
Geometrically, on a hemisphere with radius $R$ the above data describes 
a circle of radius 
$R_{\text{c}}$ that is embedded in the plane perpendicular 
to $\vec e_{\text{r}_{\text{c}}}$, i.e.\ it is i.e.\ rotated by $\psi_{\text{c}}$ to the east and inclined by $\gamma_{\text{c}}$ towards the sky and located at a distance
\begin{equation}\label{DinRc}
D=R\cos\theta_\rho
\end{equation}
from the `home' position.  Hence, the intersection of the plane at the above given distance with the hemisphere 
yields the circle. 
The vector $\vec r_{\text{c}}$ from the `home' position 
to the center of the circle is given by
\begin{equation}\label{vecrc}
\vec r_{\text{c}}=D\vec e_{\text{r}_{\text{c}}}
\pnt
\end{equation}

The orientation of the circle is fixed by applying in the case $\sigma=+1$
the right hand rule along the rotated $z$-axis of the NED coordinate system, i.e.\ along $-\vec r_{\text{c}}$. Hence, circles 
with $\sigma=+1$ and $\sigma=-1$ are oriented
clockwise and counterclockwise when observed from above, respectively.
In this convention, the orientation yields the sign for the roll angle
of an upright-oriented aircraft\footnote{Here, upright-oriented means that the pojection of the down direction of the aircraft-fixed coordinate system 
onto the down direction of the NED coordinate system is non-negative.}  
following the circular path in accord with with its orientation.

An oriented circle segment of a small circle on a hemisphere of radius $R$ can be uniquely specified by (half of) the opening angle $\theta_\rho$ defining the circle radius 
via \eqref{Rcercdef}, four angles, given as before by the 
elevation $\gamma_{\text{c}}$ and azimuthal angle $\psi_{\text{c}}$ 
as well as $0<\phi_i<2\pi$, $i=1,2$. 
The angles $\phi_1$ and $\phi_2$ determine the 
start- and endpoint of the circle segment around $\vec e_{\text{r}_{\text{c}}}$.
They are the angles of polar coordinates in a coordinate system on the plane 
with normal vector \eqref{Rcercdef}. For angles that increase in the 
direction determined by the right hand rule applied to 
$-\vec e_{\text{r}_{\text{c}}}$,
the orientation is then given by $\sigma=\sgn(\phi_2-\phi_1)$.
Alternatively, if $\phi_2-\phi_1\neq n\pi$, $n\in\mathds{Z}$,
the oriented circle segment may be defined by substituting in \eqref{Rcercdef} 
the unit vector $\vec e_{r_{\text{c}}}$ by an ordered set of two unitvectors $\vec e_{\text{t},i}=\vec e_{\text{t}}(\phi_i)$, $i=1,2$ that are the tangent vectors at the start and end point, 
respectively.
The unit normal vector  $\vec e_{\text{r}_{\text{c}}}$ pointing 
into the `upper' hemisphere towards the center of the circle segment
is then obtained as
\begin{equation}
\vec e_{\text{r}_{\text{c}}}=-\frac{\vec e_{\text{t},1}\times\vec e_{\text{t},2}}{\sin(\phi_2-\phi_1)}
\pnt
\end{equation}

At a generic position $\vec r(\phi)$ on the circle segment parameterized by
$\phi_{\text{min}}\le\phi\le\phi_{\text{max}}$, $\phi_{\text{min}}=\min(\phi_1,\phi_2)$,
 $\phi_{\text{max}}=\max(\phi_1,\phi_2)$,
the tangent space of the sphere, which is a plane, contains the tangent 
vector $\vec e_{\text{t}}(\phi)$ of the
curve. The direction normal to the sphere and the tangent plane
are spanned by the unit vectors
\begin{equation}
\vec e_{\text{r}}(\phi)=\frac{\vec r(\phi)}{|\vec r(\phi)|}\col\qquad
\vec e_{\text{p},1}(\phi)=\vec e_{\text{t}}(\phi)\col\qquad
\vec e_{\text{p},2}(\phi)=\vec e_{\text{r}}(\phi)\times\vec e_{\text{p},1}(\phi)
\pnt
\end{equation} 
The decomposition of the curvature $\kappa$ of a curve on the sphere
into the geodesic curvature $\kappa_{\text{g}}$ associated with the projection
of the curve onto the tangent plane of the sphere and the
normal curvature $\kappa_{\text{n}}$
read
\begin{equation}\label{kappa}
\frac{\partial_\phi\vec e_{\text{t}}(\phi)}{|\partial_\phi\vec r(\phi)|}
=\kappa(\phi)\vec e_{\text{n}}(\phi)
=\kappa_{\text{g}}(\phi)\vec e_{\text{p},2}(\phi)+\kappa_{\text{n}}\vec e_{\text{r}}(\phi)
\col\qquad 
\kappa_{\text{n}}=-\frac{1}{R}
\col
\end{equation}
where $\vec e_{\text{n}}(\phi)$ is the normal vector of the
curve.\footnote{Here, $\vec e_{\text{n}}(\phi)$ is defined such that $\kappa(\phi)\ge0$ and therefore is not the outer normal vector but points into the direction into which the curve deviates from a straight line, 
which for a circle is the interior.
Hence, \eqref{kappa} differs by a sign from \eqref{kappal} which is formulated for the outer normal vector.}
For a circle segment, $\kappa=\frac{1}{R_{\text{c}}}$ is constant, and the 
geodesic curvature reads with the definition of $\theta_\rho$ in \eqref{DinRc}
\begin{equation}
\kappa_{\text{g}}=\frac{\sqrt{R^2-R_{\text{c}}^2}}{RR_{\text{c}}}
=\frac{\cot\theta_\rho}{R}
\pnt
\end{equation}
In the special case $\theta_\rho=\frac{\pi}{2}$, the geodesic curvature
is zero, indicating that the resulting circle segment is a geodesic, i.e.\ a 
great circle segment. Indeed, the plane defined via \eqref{DinRc} and \eqref{vecrc} then
contains the center point of the hemisphere since $D=0$ and thus yields
a great circle segment.

The oriented figure-eight pattern 
shown in figure \ref{fig:flightcircleeight} can be constructed from four oriented circle segments, of which two are great circle segments
and two are small circle segments.
The two great circle segments are of equal length and intersect each other at half of their length. The intersection is called the crossing point of the figure-eight pattern. Each pair of neighboured endpoints is connected via 
a small circle. If the plane defining this small circle contains the respective two endpoints of the great circles and is spanned by the tangent vectors 
of the great circles at these points, the piecewise defined curve is $C^1$ 
at the connections.

In the NED coordinate system, 
the figure-eight pattern is most easily constructed 
at the pole of the `upper' hemisphere with an orientation from east to west.
The location of the crossing point and vectors pointing towards
the centers of the eastern ($\sigma_{\text{e}}=+1$) and western ($\sigma_{\text{e}}=-1$) turning circles according to \eqref{Rcercdef} 
are then given by
\begin{equation}\label{r0}
\vec r_0 =R\begin{pmatrix} 0 \\ 0 \\ -1 \end{pmatrix}
\col\qquad
\vec e_{\text{r}_{\text{c}},\sigma_{\text{e}}}
=\begin{pmatrix} 0 \\ \sigma_{\text{e}}\sin\theta_{\text{c}} \\ -\cos\theta_{\text{c}} \end{pmatrix}
\col
\end{equation}
where $\theta_{\text{c}}\ge\theta_\rho$ is required in order to avoid multiple
self intersections. For an orientation $\sigma=+1$ of the figure-eight pattern,
the eastern and western turning circles are
oriented clockwise and counter clockwise when observed from above, respectively. Their orientations are then given by $\sigma_{\text{e}}$.

The two geodesic segments that intersect in $\vec r_0$ at an angle 
$0<\chi_0\le\pi$ are parameterized as
\begin{equation}
\vec r_{\text{g}\pm}(\theta)
=R\begin{pmatrix} \pm\sin\theta\sin\tfrac{\chi_0}{2} \\ \sin\theta\cos\tfrac{\chi_0}{2} \\  -\cos\theta
\end{pmatrix}
\col\qquad
-\theta_{\text{t}}\le\theta\le\theta_{\text{t}}
\col
\end{equation}
where $\theta_{\text{t}}$ is the polar angle at the two eastern endpoints of 
the two geodesic segments. At these points and their two western counterparts, 
the transgressions between the segments occur, and we will hence from now 
on denote them as transgression points. 
The projection of $\vec r_{\text{g}\pm}(\theta_{\text{t}})$ 
onto the direction $\vec e_{\text{r}_\text{c},+}$ of the eastern turning circle 
center yields
\begin{equation}\label{projrel}
\frac{\vec r_{\text{g},\pm}(\theta_{\text{t}})\cdot\vec e_{\text{r}_\text{c},+}}{R}
=(\sin\theta_{\text{t}}\sin\theta_{\text{c}}\cos\tfrac{\chi_0}{2}
+\cos\theta_{\text{t}}\cos\theta_{\text{c}})
=\cos\theta_\rho
\pnt
\end{equation}

The unit tangential vectors at the 
north-eastern, north-western, south-eastern, south-western transgression points
labeled by $(\sigma_{\text{n}},\sigma_{\text{e}})=(+,+),(+,-),(-,+),(-,-)$
and the two required cross products read
\begin{equation}
\vec e_{\text{t},\sigma_{\text{n}},\sigma_{\text{e}}}
=\begin{pmatrix} \cos\theta_{\text{t}}\sin\tfrac{\chi_0}{2} \\ \sigma_{\text{n}}\sigma_{\text{e}}\cos\theta_{\text{t}}\cos\tfrac{\chi_0}{2} \\ \sigma_{\text{n}}\sin\theta_{\text{t}}
\end{pmatrix}
\col\quad
\sigma_{\text{e}}\,\vec e_{\text{t},+,\sigma_{\text{e}}}\times\vec e_{\text{t},-,\sigma_{\text{e}}}
=2\cos\theta_{\text{t}}\sin\tfrac{\chi_0}{2}
\begin{pmatrix}  0 \\ \sigma_{\text{e}}\sin\theta_{\text{t}} \\ -\cos\theta_{\text{t}}\cos\tfrac{\chi_0}{2}
\end{pmatrix}
\pnt
\end{equation}
The latter expression has to be parallel to $\vec e_{\text{r}_{\text{c}},\sigma_{\text{e}}}$ in \eqref{r0}, and this determines the polar angle of the eastern center point direction as
\begin{equation}\label{dirrel}
\tan\theta_{\text{c}}=\frac{\tan\theta_{\text{t}}}{\cos\tfrac{\chi_0}{2}}
\pnt
\end{equation}
The relations \eqref{projrel} and \eqref{dirrel} determine $\theta_{\text{t}}$ and 
$\chi_0$ in terms of the given angles $\theta_{\text{c}}$ and $\theta_\rho$ as
\begin{equation}
\cos\theta_{\text{t}}=\frac{\cos\theta_{\text{c}}}{\cos\theta_\rho}\col\qquad
\cos\tfrac{\chi_0}{2}
=\frac{\sqrt{\sin^2\theta_{\text{c}}-\sin^2\theta_\rho}}{\sin\theta_{\text{c}}}
 \pnt
\end{equation}

The position vectors of the four transgression points and their tangential vectors determining the orientation of the segments and outer normal vectors 
are then expressed as functions of $\theta_{\text{c}}$ and $\theta_\rho$ as
\begin{equation}\label{rgpm}
\begin{aligned}
\vec r_{\text{g},\sigma_{\text{n}},\sigma_{\text{e}}}
&=
\frac{R}{\cos\theta_\rho\sin\theta_{\text{c}}}
\begin{pmatrix} 
\sigma_{\text{n}} \sin\theta_\rho\sqrt{\sin^2\theta_{\text{c}}-\sin^2\theta_\rho} \\ 
\sigma_{\text{e}}(\sin^2\theta_{\text{c}}-\sin^2\theta_\rho)\\ 
-\sin\theta_{\text{c}}\cos\theta_{\text{c}}
\end{pmatrix}
\col\\
\vec e_{\text{t},\sigma_{\text{n}},\sigma_{\text{e}}}
&=
\frac{1}{\cos\theta_\rho}
\begin{pmatrix} 
\sin\theta_\rho\cot\theta_{\text{c}} \\
\sigma_{\text{n}}\sigma_{\text{e}}\cot\theta_{\text{c}}\sqrt{\sin^2\theta_{\text{c}}-\sin^2\theta_\rho} \\
\sigma_{\text{n}}\sqrt{\sin^2\theta_{\text{c}}-\sin^2\theta_\rho}
\end{pmatrix}
\col\\
\vec e_{\text{n},\sigma_{\text{n}},\sigma_{\text{e}}}
&=\sigma_{\text{e}}\,\vec e_{\text{r}_{\text{c}},\sigma_{\text{e}}}\times\vec e_{\text{t},\sigma_{\text{n}},\sigma_{\text{e}}}
=\frac{1}{\cos\theta_\rho\sin\theta_{\text{c}}}\begin{pmatrix} \sigma_{\text{n}}\sqrt{\sin^2\theta_{\text{c}}-\sin^2\theta_\rho} \\ -\sigma_{\text{e}}\sin\theta_\rho\cos^2\theta_{\text{c}} \\ -\sin\theta_\rho\sin\theta_{\text{c}}\cos\theta_{\text{c}} \end{pmatrix}
\pnt
\end{aligned}
\end{equation}

The angle $\phi_{\text{c}}$ swept by 
each small circle segment is determined as
\begin{equation}
\begin{aligned}
\sin\tfrac{\phi_{\text{c}}}{2}
=\sqrt{\frac{1+\vec e_{\text{t},+,\sigma_{\text{e}}}\cdot\vec e_{\text{t},-,\sigma_{\text{e}}}}{2}}
=\frac{\tan\theta_\rho}{\tan\theta_{\text{c}}}
\pnt
\end{aligned}
\end{equation}

According to \eqref{DinRc} and\eqref{vecrc}, the two center points of the 
turning circles and apices of the figure-eight pattern are located at
\begin{equation}
\vec r_{\text{c},\sigma_{\text{e}}}=R\cos\theta_\rho\begin{pmatrix} 0 \\ \sigma_{\text{e}}\sin\theta_{\text{c}} \\ -\cos\theta_{\text{c}} \end{pmatrix}
\col\qquad
\hat r_{\sigma_{\text{e}}}
=\vec r_{\text{c},\sigma_{\text{e}}}+R\sin\theta_\rho\begin{pmatrix} 0 \\ \sigma_{\text{e}}\cos\theta_{\text{c}} \\ \sin\theta_{\text{c}} \end{pmatrix}
=R\begin{pmatrix} 0 \\ \sigma_{\text{e}}\sin\hat\theta \\ \cos\hat\theta \end{pmatrix}
\col
\end{equation}
where (half of) the maximum swept polar angle
is given by
\begin{equation}
\hat\theta=\theta_{\text{c}}+\theta_\rho\pnt
\end{equation}
Clearly, to stay on the `upper' hemisphere, one has to choose $\theta_{\text{c}}$ and $\theta_\rho$ such that $\hat\theta\le\frac{\pi}{2}$.

The figure-eight pattern with a generic azimuthal and elevation angle 
such that the crossing point $\vec r_0$ in \eqref{r0} is 
transformed to $\vec r_0=(\sin\theta_0\cos\psi_0,\sin\theta_0\sin\psi_0,-\cos\theta_0)^{\text{t}}$ can be obtained by applying a rotation $R(\theta_0,\psi_0)$.
The orientation of the figure-eight pattern should always be horizontal, i.e.\ the two turning points should always have the same $z$-coordinate. This can be achieved by first performing an active rotation with angle $-\theta_0$ around the $y$-axis. Then, one performs an active rotation with angle
$\psi_0$ around the $z$-axis.

The rotation matrix that combines two such active rotations 
but with angles given by $-\theta$ and $\psi$, reads
\begin{equation}
R(\theta,\psi)
=(\vec e_{\theta}, \vec e_{\psi}, -\vec e_{\text{r}})
\col
\end{equation}
where vectors in the directions of the north, east, down 
axes are mapped to
\begin{equation}\label{es}
\vec e_{\theta}
=\begin{pmatrix}
\cos\theta\,\vec e_1 \\
\sin\theta
\end{pmatrix}
\col\qquad
\vec e_{\psi}
=\begin{pmatrix}
\vec e_2 \\
0
\end{pmatrix}
\col\qquad
\vec e_{\text{r}}
=\begin{pmatrix}
\sin\theta\,\vec e_1 \\
-\cos\theta
\end{pmatrix}
\col
\end{equation}
respectively. 
The basis vectors in the lateral (north-east) subspace are given by
\begin{equation}
\vec e_1
=\begin{pmatrix}
\cos\psi \\
\sin\psi \\
\end{pmatrix}
\col\qquad
\vec e_2
=\begin{pmatrix}
-\sin\psi \\
\cos\psi
\end{pmatrix}
\pnt
\end{equation}
With $\theta_0=\frac{\pi}{2}-\gamma_0$, the oriented figure-eight is hence specified by its
\begin{equation}
\begin{aligned}
&\text{shape: }&&0<\theta_{\text{c}}<\frac{\pi}{2}\col&\qquad
&0<\theta_\rho\le\theta_{\text{c}}\col&\qquad
\theta_{\text{c}}+\theta_\rho\le\frac{\pi}{2}\col\\
&\text{attitude: }&&0\le\gamma_0\le\frac{\pi}{2}\col&\qquad
&0\le\psi_0\le 2\pi\col\\
&\text{orientation: }&&\sigma=\pm1
\pnt
\end{aligned}
\end{equation}

\subsection{Navigation and control strategy}

The navigation of the aircraft is divided into lateral (NE subspace)
navigation and speed-height control. The navigation controller determines the 
demanded lateral acceleration of the aircraft and computes the demanded
roll angle from it. Furthermore, the demanded height is transferred to
the speed-height-controller. This control strategy requires the decomposition
of the desired flight path into lateral and height components.

The small circle defined via \eqref{Rcercdef} can be parameterized in 
terms of the time-dependent angle $0\le\phi(t)<2\pi$. For $0<\theta_c<\frac{\pi}{2}$,
the vector $\vec e_1$ shall point from the lateral projection of the center point of the circle to that of the initial point on the circle where $\phi=0$.\footnote{For $\theta_c=0$, we choose $\vec e_1$ to point into the northern direction.}
This parameterization together with the resulting velocity 
reads
\begin{equation}
\begin{aligned}
\vec r(\phi)
&=R(\cos\theta_\rho\,\vec e_{\text{r}_{\text{c}}}
+\sin\theta_\rho(\cos\phi\,\vec e_{\theta_{\text{c}}}
+\sigma\sin\phi\,\vec e_{\psi_{\text{c}}}))
\col\\
\dot{\vec r}(\phi)
&=-v_{\text{c}}(\sin\phi\,\vec e_{\theta_{\text{c}}}
-\sigma\cos\phi\,\vec e_{\psi_{\text{c}}})
\col\qquad 
v_{\text{c}}=R\dot\phi\sin\theta_\rho
\col
\end{aligned}
\end{equation}
where the occuring unit vectors are the ones of \eqref{es} with angles
$\theta_{\text{c}}$, $\psi_{\text{c}}$. 
This parameterization decomposes into the lateral and height direction as
\begin{equation}
\begin{aligned}\label{latheightdec}
\vec r_{\text{l}}(\phi)
&=R(\cos\theta_\rho\sin\theta_{\text{c}}\,\vec e_1
+\sin\theta_\rho(\cos\phi\cos\theta_{\text{c}}\,\vec e_1
+\sigma\sin\phi\,\vec e_2))
\col\\
z(\phi)
&=-R(\cos\theta_\rho\cos\theta_{\text{c}}
-\cos\phi\sin\theta_\rho\sin\theta_{\text{c}})
\col\\
\dot{\vec r}_{\text{l}}(\phi)
&=-v_{\text{c}}(\sin\phi\cos\theta_{\text{c}}\,\vec e_1
-\sigma\cos\phi\,\vec e_2)
\col\\
\dot z(\phi)
&=-v_{\text{c}}\sin\phi\sin\theta_{\text{c}}
\pnt
\end{aligned}
\end{equation}

The unit tangent and outer normal vector in the lateral plane obtained from the above parameterization read
\begin{equation}
\begin{aligned}\label{etlenl}
\vec e_{\text{tl}}(\phi)
&=
-\frac{\sin\phi\cos\theta_{\text{c}}\,\vec e_1-\sigma\cos\phi\,\vec e_2}{\sqrt{\cos^2\phi+\sin^2\phi\cos^2\theta_{\text{c}}}}
\col\qquad
\vec e_{\text{nl}}(\phi)
=\frac{\cos\phi\,\vec e_1+\sigma\sin\phi\cos\theta_{\text{c}}\,\vec e_2}{\sqrt{\cos^2\phi+\sin^2\phi\cos^2\theta_{\text{c}}}}
\pnt
\end{aligned}
\end{equation}
They determine the lateral curvature $\kappa_{\text{l}}$ as
\begin{equation}
\begin{aligned}\label{kappal}
\frac{\partial_\phi\vec e_{\text{tl}}(\phi)}{|\partial_\phi\vec r_{\text{l}}(\phi)|}
=-\kappa_{\text{l}}(\phi)\vec e_{\text{nl}}(\phi)\col\qquad
\kappa_{\text{l}}(\phi)=\frac{\cos\theta_{\text{c}}}{R\sin\theta_\rho\sqrt{\cos^2\phi+\sin^2\phi\cos^2\theta_{\text{c}}}^3}
\pnt
\end{aligned}
\end{equation}
It enters the expression for the centripetal acceleration that is 
required in order to stay on the given curve
\begin{equation}\label{al}
a_{\text{l}}(\phi)=v_{\text{l}}(\phi)^2\kappa_{\text{l}}(\phi)
=\frac{v_{\text{c}}^2}{R\sin\theta_\rho}\frac{\cos\theta_{\text{c}}}{\sqrt{\cos^2\phi+\sin^2\phi\cos^2\theta_{\text{c}}}}
\col
\end{equation}
where the lateral velocity is determined from \eqref{latheightdec} as 
\begin{equation}
v_{\text{l}}(\phi)=|\dot{\vec r}_{\text{l}}(\phi)|
=v_{\text{c}}\sqrt{\cos^2\phi+\sin^2\phi\cos^2\theta_{\text{c}}}
\pnt
\end{equation}

In order to follow the desired flight path, 
the controller has to compare the measured with the desired position 
and attitude. The distance of the measured position from the desired 
flight path and also the deviation of the 
flight direction from the tangential direction at the corresponding location 
on the flight path have to be determined.
 Let  $\vec r_{\text{a}}$ and  $\vec v_{\text{a}}$ be the
measured position and velocity of the aircraft, respectively.
The position vector $\vec r_{\text{p}}$ of the closest point on the circle 
is given by
\begin{equation}
\vec r_{\text{p}}=D\vec e_{\text{r}_{\text{c}}}+R\sin\theta_\rho\,\vec e_{\text{np}}
\col
\end{equation}
where the tangent vector and outer normal vector at the position $\vec r_{\text{p}}$ 
on the circle read
\begin{equation}
\vec e_{\text{tp}}=\sigma\frac{\vec r_{\text{a}}\times\vec e_{\text{r}_{\text{c}}}}{|\vec r_{\text{a}}\times\vec e_{\text{r}_{\text{c}}}|}
\col\qquad
\vec e_{\text{np}}=\sigma\,\vec e_{\text{r}_{\text{c}}}\times\vec e_{\text{tp}}
=\frac{\vec r_{\text{a}}-(\vec r_{\text{a}}\cdot \vec e_{\text{r}_{\text{c}}})\vec e_{\text{r}_{\text{c}}}}{\sqrt{r_{\text{a}}^2-(\vec r_{\text{a}}\cdot \vec e_{\text{r}_{\text{c}}})^2}}
\col
\end{equation}
and they obey $\vec e_{\text{tp}}\times\vec e_{\text{np}}=\sigma\,\vec e_{\text{r}_{\text{c}}}$.
The decomposition into lateral direction and height requires
their normalized lateral projections and the outer normal vector
\begin{equation}
\begin{aligned}
\vec e_{\text{tpl}}
=\frac{\vec e_{\text{tp}}-e_{\text{tp},3}\vec e_3}{\sqrt{1-e_{\text{tp},3}^2}}
\col\qquad
\vec e_{\text{rpl}}=\frac{\vec e_{\text{np}}-e_{\text{np},3}\vec e_3}{\sqrt{1-e_{\text{np},3}^2}}
\col\qquad
\vec e_{\text{npl}}
=\sigma\frac{\vec e_{\text{tp}}\times\vec e_3}{\sqrt{1-e_{\text{tp},3}^2}}
\pnt
\end{aligned}
\end{equation}
While $\vec e_{\text{npl}}$ has been constructed such that 
$\vec e_{\text{tpl}}\cdot \vec e_{\text{npl}}=0$, this is not the case for
$\vec e_{\text{rpl}}$. In fact, unlike for the inclined circle, for an ellipse
which is its lateral projection,
the lateral radial and normal vector are not parallel at a generic position. 
Their inner product reads
\begin{equation}
\begin{aligned}
\vec e_{\text{rpl}}\cdot\vec e_{\text{npl}}
=\frac{|e_{\text{r}_{\text{c}},3}|}{\sqrt{(1-e_{\text{tp},3}^2)(1-e_{\text{np},3}^2)}}
\pnt
\end{aligned}
\end{equation}

The result for the lateral acceleration \eqref{al} can be obtained 
even without knowing the value of $\phi$ at the location $\vec r_{\text{p}}$.
From \eqref{etlenl} it follows that 
\begin{equation}\label{alfinal}
a_{\text{l}}=\frac{v_{\text{c}}^2}{R\sin\theta_\rho}
\sqrt{\cos^2\theta_{\text{c}}\,(\vec e_1\cdot\vec e_{\text{npl}})^2+(\vec e_2\cdot\vec e_{\text{npl}})^2}
\pnt
\end{equation}

The deviation of the measured from the desired position,
its lateral projection and its component in normal direction
read
\begin{equation}
\begin{aligned}\label{deltarh}
\vec r_{\text{pa}}&=\vec r_{\text{a}}-\vec r_{\text{p}}
\col\quad
\vec r_{\text{pal}}
=\vec r_{\text{pa}}-r_{\text{pa},3}\vec e_3
\col\quad
\delta r_{\text{l}}= \sgn(\vec r_{\text{pal}}\cdot\vec e_{\text{npl}})|\vec r_{\text{pal}}|
\col\quad
h=-r_{\text{p},3}
\col
\end{aligned}
\end{equation}
where we have also indicated the demanded height that is transferred
to the speed-height controller.
Note that $\delta r_{\text{l}}$ is not the distance of 
the aircraft's position from the ellipse, since the lateral projection of the
nearest point on the circle does in general not yield the nearest point 
on the resulting ellipse. It can nevertheless be used as a measure of 
the deviation from the desired flight path.

The lateral projection of the velocity vector and its component in 
normal direction read
\begin{equation}
\begin{aligned}
\vec v_{\text{pal}}
&=\vec v_{\text{a}}-v_{\text{a},3}\vec e_3
\col\qquad
\delta v_{\text{l}}
=\vec v_{\text{pal}}\cdot\vec e_{\text{npl}}
\pnt
\end{aligned}
\end{equation}

The acceleration that has to be provided by the appropriate roll angle reads
\begin{equation}\label{alateral}
a=a_{\text{l}}+ K_r\delta r_{\text{l}} + K_v\delta v_{\text{l}} 
\col
\end{equation}
where $K_r$ and $K_v$ are positive gains of the 
proportional and differential feedback that yield the PD-controller, respectively. 

The roll angle $\Phi$ is determined from the equilibrium of the centripetal, lift and gravitational forces acting 
on the aircraft. Considering a pitch angle $\Theta$, the
result reads
\begin{equation}\label{Phi}
\Phi=\sigma\arctan\tfrac{a\cos\Theta}{g}
\col\qquad -\frac{\pi}{2}\le\Phi\le\frac{\pi}{2}
\pnt
\end{equation}
The tether force \eqref{tetherforce} is not considered in the calculation
of $\Phi$.

\subsection{Code for fixed-wing aircraft}

For a fixed-wing aircraft, the vehicle specific code of the {\tt ardupilot} 
autopilot  is denoted as {\tt ArduPlane}. In the following, it is
described how flight modes are embedded in this code.
The new flight modes for tethered flight along
rotated and inclined circles and figure-eight patterns are 
called {\tt LOITER\_3D} and {\tt EIGHT\_SPHERE}, respectively.

\begin{figure}[t]
\centering
\includegraphics[trim = 27mm 103mm 23mm 31mm, clip]{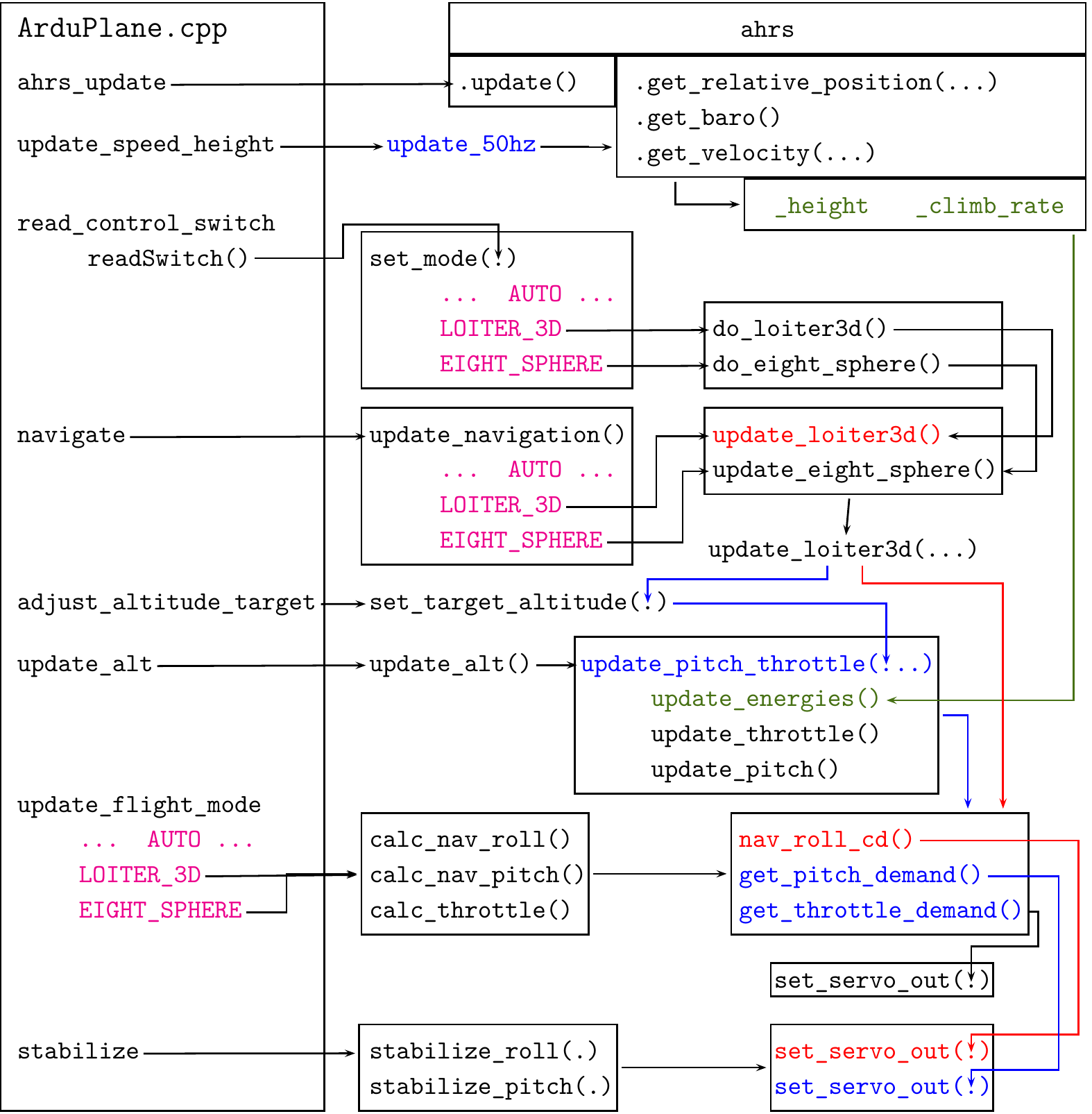}%
\caption{Embedding of the new flight modes \textcolor{magenta}{\tt LOITER\_3D} and \textcolor{magenta}{\tt EIGH\_SPHERE}
into the {\tt ardupilot} code for fixed-wing aircraft ({\tt ArduPlane}).
Speed and height are controlled by the \textcolor{blue}{total energy control system (TECS)}. Lateral navigation is provided by an \textcolor{red}{L1-controller}
 \cite{park04} in case of straight flight paths or a \textcolor{red}{PD-controller} in case of
circles or circle segments.
\label{fig:tasks}}
\end{figure}

The main file of the code for fixed-wing aircraft is {\tt ArduPilot.cpp}
with its header {\tt Plane.h}. It contains
the list of tasks called {\tt scheduler\_tasks}.
In the main loop function {\tt loop()}, these tasks are executed after a sample has been obtained from the sensors. It runs at a rate
{\tt \_loop\_rate\_hz}=400Hz, i.e. every {\tt loop\_us}=2500$\mu\text{s}$. The tasks run at individual rates, i.e.\ not each task is due to run
in each main loop cycle. The tasks of immediate relevance for navigation are shown in figure \ref{fig:tasks} in the box labeled {\tt ArduPlane.cpp}. 
The list is causally ordered, and this order does not necessarily
coincide with the temporal order in which the individual tasks are 
executed by the scheduler. 

First, {\tt ahrs\_update} calls the {\tt update()} method from the 
attitude and height reference system {\tt AHRS} class. It uses an extended Kalman filter (EKF) to fuse the sensor data in order to determine the state of the aircraft, e.g\ its position and attitude. The task {\tt update\_speed\_height}
then reads that data with respective methods of the {\tt ahrs} object and
stores the calculated  height and climb rate  
in the variables {\tt \_height} and  {\tt \_climb\_rate}, respectively.
The task {\tt read\_control\_switch} selects the flight mode
according to the positions of the control switches of the  
RC transmitter or GCS.
In figure \ref{fig:tasks}, some of these modes are shown and highlighted 
in magenta. The mode {\tt AUTO} navigates along a polygon 
given in terms of corner points (waypoints).
The modes for tethered flight are 
{\tt LOITER\_3D} and {\tt EIGHT\_SPHERE}. They implement the flight along 
given rotated and inclined versions of the circle and figure-eight pattern
shown in figure \ref{fig:flightcircleeight}, respectively.
The selected flight mode is initialized
when the corresponding control switch is activated. 
For the modes {\tt LOITER\_3D} and {\tt EIGHT\_SPHERE} this
happens by calling {\tt do\_loiter3d()} and {\tt do\_eight\_sphere()}, respectively. Then, the task {\tt navigate} 
calls the virtual function {\tt update\_loiter3d()} or {\tt update\_eight\_sphere()}. Via dynamic binding they call their counterparts defined in the selected 
navigation controller class which is {\tt A\_L1\_control}.
In the {\tt LOITER\_3D} case, the called function is
{\tt update\_loiter3d(...)} that takes certain parameters
as its argument, calculates the demanded height given in \eqref{deltarh} and employs a PD-controller to calculate the demanded lateral acceleration given in \eqref{alateral}. In the {\tt EIGHT\_SPHERE} case, {\tt update\_eight\_sphere()} selects the current segment of the figure-eight pattern in dependence of the current position of the aircraft. Then, it calls {\tt update\_loiter3d(...)} 
of the {\tt A\_L1\_control} class with the parameters of the selected segment.
The demanded height \eqref{deltarh} is then
transferred via the function {\tt set\_target\_alitude} of the
task {\tt adjust\_altitude\_target} to the total energy control system (TECS)
that is called from the task {\tt update\_alt}. It  
calculates the desired pitch and throttle from the demanded height and 
the measured {\tt \_height} and {\tt \_climb\_rate}.
The task {\tt update\_flight\_mode} then calculates the roll angle
from the accelerations according to \eqref{Phi} in {\tt nav\_roll\_cd()} 
and simply reads out the pitch and throttle demand 
that were determined by the TECS. Moreover, it transfers the throttle demand
to the servos via the function {\tt set\_servo\_out}. 
The task {\tt stabilize} then also transfers the required roll and pitch 
angle to the servos.

\section{Simulation}

We have tested the modes for tethered flight in simulations. This is 
recommended in order to debug the code and to reduce the risk of malfunction
in real tests. The autopilot {\tt ardupilot} admits compilation for a
SITL environment. It uses the flight dynamics model {\tt JSBsim} with
{\tt MAVProxy} as GCS. The models of aircraft are already included
in the simulation software. The 
default model is for the Rascal 110, which is used for the tests of the 
tethered flight modes. The various coefficients and characteristics of the 
aircraft model are specified in several files via the extended markup language (XML). In particular, it is possible to add external forces acting on the aircraft in specified directions at specified positions. A model for the tether force can be implemented in this way. We have added it to the aircraft model as {\tt <external\_reactions>} that 
acts at the center of mass of the aircraft in the direction of the unit vector pointing from that point to the `home' location. The strength of the
force depends on the distance from `home' as given in \eqref{tetherforce}.
So far, we have not implemented a model of our test platform, i.e. of the
modified Easy Star II. But even without 
such a model the SITL environment was invaluable for debugging the code 
and for ensuring that our aircraft can be safely operated in the flight tests
described below.

\section{Flight tests and data analysis}

In this section, we analyze the data from our flight tests. 
The tests start with a flight of a horizontal circle
at fixed altitude without
tether in order to calibrate
the airspeed sensor and to determine the wind direction and speed.
Then, the aircraft is ready for testing the tethered flight modes.

The autopilot
stores the system status and state as well as environmental data
into several log files. One of these contains the data as lines
of comma separated values (CSV) in temporal order
of which the first entry of each line
is an identifier of the source of information.
For each identifier the format as well as type of its data is specified
in the preamble.
For instance, the identifiers {\tt NKF1} to {\tt NKF4} contain 
information for the first extended Kalman filter (EKF) instance
\cite{EKF2}.
The following analysis relies mainly on the {\tt NKF1} line that contains 
attitude, position, ground speed data of the aircraft and on the 
{\tt AHRS} line that contains the measured airspeed. Moreover, the {\tt CTUN}
line is used to retrieve information about the throttle.
Thereby, data from different sources such as e.g.\ the position from {\tt NKF1}
and the airspeed from {\tt AHRS} has to be combined. 
This data is logged at different time instances and with different
frequencies. We searched for 
that instance of the desired data logged with the higher frequency ({\tt NKF1})
that occurs immediately before an instance of the desired data logged with 
the lower frequency ({\tt AHRS} or {\tt CTUN}) and combined them. 
Since they do not occur at coincident
times, we checked that their time differences are small enough
to be negligible. 

\subsection{Calibration of the airspeed sensor}
\label{sec:airspeedcalibration}

The airspeed sensor measures the airspeed via a Pitot tube that is mounted  
at the nose of the aircraft as shown in figure \ref{fig:components}. 
It has to be calibrated in order to 
provide reliable results for the airspeed. 
\begin{figure}[t]
\centering
\begin{subfigure}{0.5\textwidth}
\includegraphics[width=\textwidth]{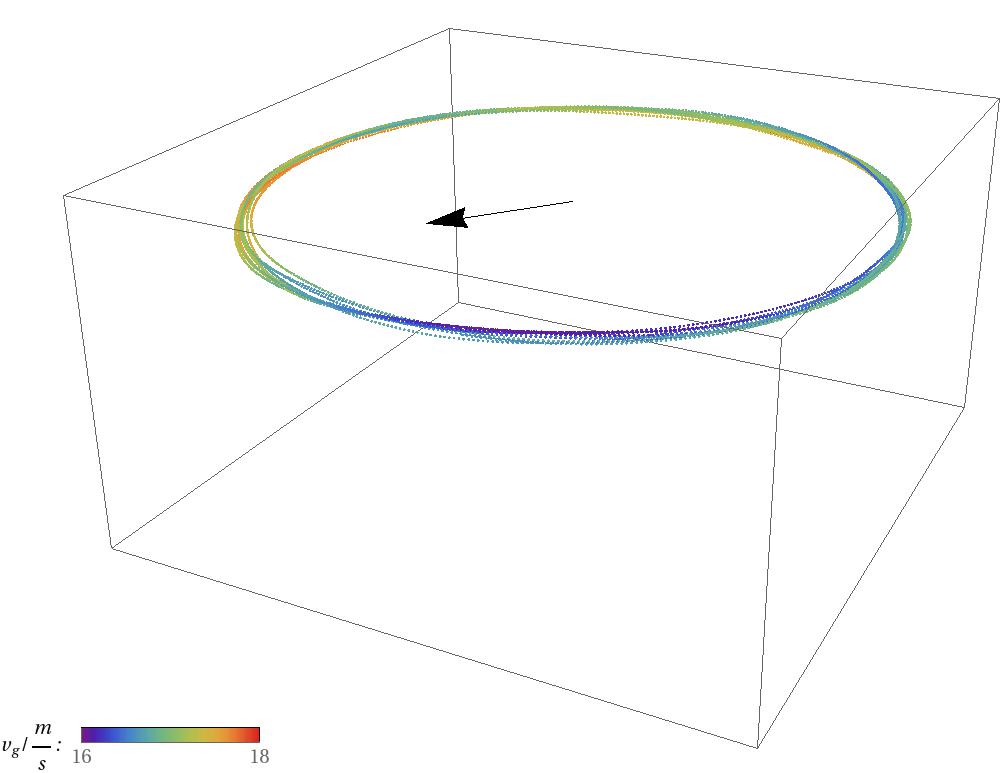}%
\end{subfigure}%
\begin{subfigure}{0.5\textwidth}
\includegraphics[width=\textwidth]{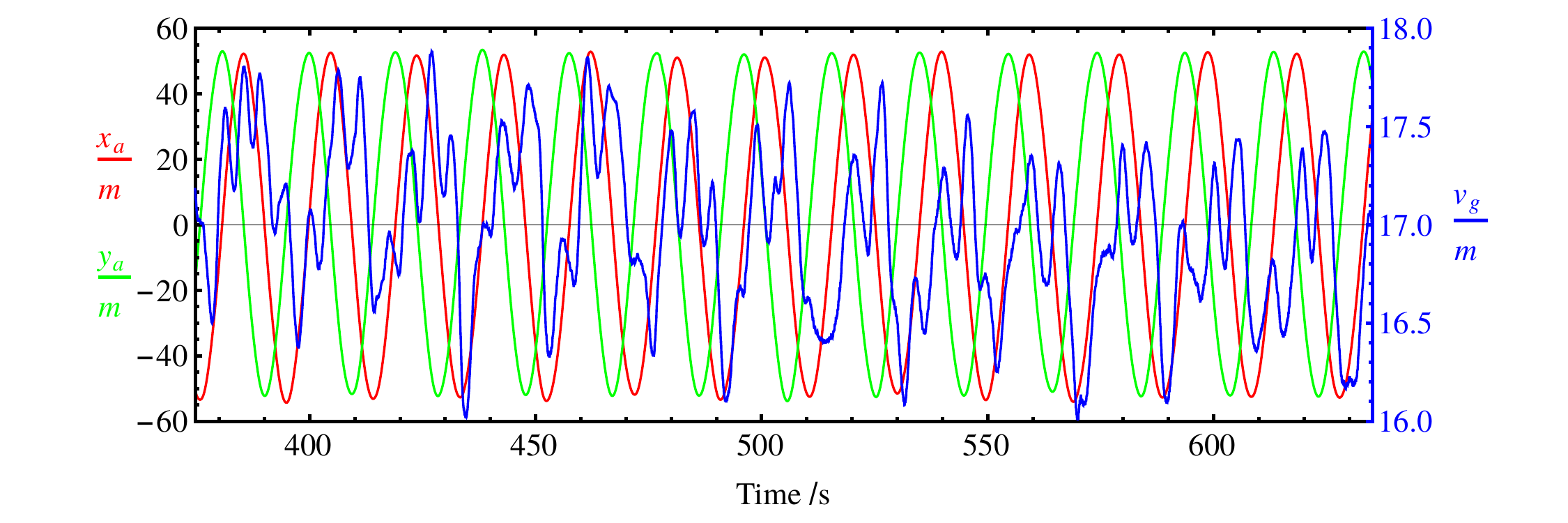}\\%
\includegraphics[width=\textwidth]{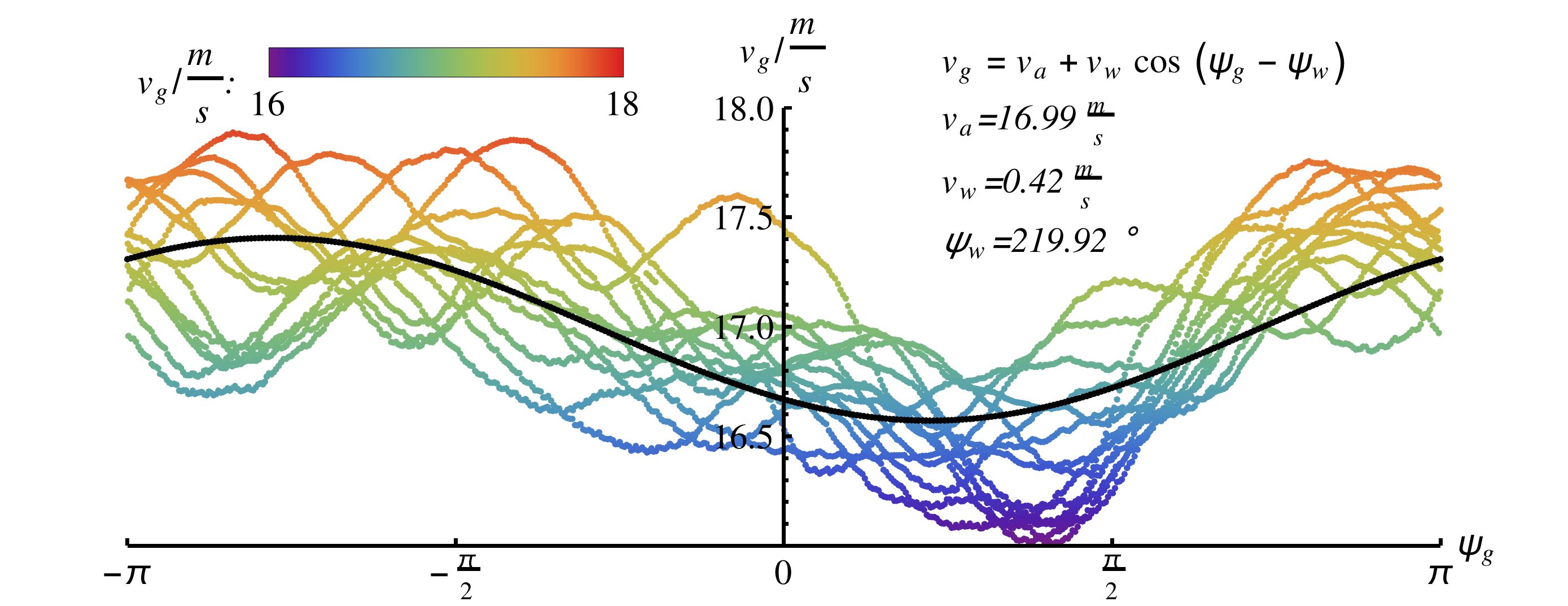}%
\end{subfigure}%
\caption{\label{fig:cARSP}
Circular path used for the calibration of the airspeed sensor. The ground speed $v_{\text{g}}$ is encoded by the color and the determined wind direction is shown. The time-dependence of the measured lateral position and ground speed velocity as well as the fit to the data is shown on the rhs.}
\end{figure}
To this purpose the 
aircraft flies along a horizontal circle several times, keeping constant the airspeed $v_{\text{a}}$, where we have demand $v_{\text{a}}=17\frac{\text{m}}{\text{s}}$. Since the ground speed $v_{\text{g}}$ is determined by the Kalman filter from the measured accelerations and GPS positions, the relation
\begin{equation}
\vec v_{\text{g}}=\vec v_{\text{a}}+\vec v_{\text{w}}
\end{equation}
can be used to determine $v_{\text{a}}$ and the average 
wind speed $v_{\text{w}}$ and wind direction $\psi_{\text{w}}$ in 
the NED coordinate system
from the time series of data points. 
If $v_{\text{w}}\ll v_{\text{g}}, v_{\text{a}}$ as in the present case, $\vec v_{\text{g}}$ and $\vec v_{\text{a}}$, almost point into the same direction.
The above relation then simplifies for a
circular flight path defined via
\eqref{Rcercdef} with $\gamma=90^\circ$ 
the following relation for the lateral components
\begin{equation}\label{vgvavwrelsimp}
(v_{\text{g}}-v_{\text{a}})\begin{pmatrix}\cos\psi_{\text{g}} \\ \sin\psi_{\text{g}} \end{pmatrix}
=v_{\text{w}}\begin{pmatrix}\cos\psi_{\text{w}} \\ \sin\psi_{\text{w}}\end{pmatrix}
\col
\end{equation}
where $\psi_{\text{g}}$ is the angle corresponding to the direction of the ground speed vector. On the considered circular path, $\vec v_{\text{g}}$ and the vector pointing radially from the center of the circle to the position of the aicraft
are perpendicular. Hence, the two angles corresponding to their directions
are related as as
$\psi_{\text{g}}=\frac{\pi}{2}-\psi_{\text{a}}$, where $\psi_{\text{a}}$ is the
angle associated with the radial vector.
Contracting both sides of \eqref{vgvavwrelsimp} with
$(\cos\psi_{\text{g}}, \sin\psi_{\text{g}})^\T$ yields
the ground speed as a function of $\psi_{\text{g}}$
\begin{equation}\label{vginpsig}
v_{\text{g}}(\psi_{\text{g}})=v_{\text{a}}+v_{\text{w}}\cos(\psi_{\text{g}}-\psi_{\text{w}})
\pnt
\end{equation}
A fit of this function to the measured data determines $v_{\text{a}}$,
$v_{\text{w}}$ and $\psi_{\text{w}}$. 
In figure \ref{fig:cARSP}, the
circular flight path used for the calibration of the airspeed sensor is shown.
Its color encodes the value of $v_{\text{g}}$. Moreover, the time dependence
of the lateral position and ground speed as well as the fit of \eqref{vginpsig}
is shown. The result for the airspeed matches the desired airspeed
of $v_{\text{a}}=17\frac{\text{m}}{\text{s}}$ very well. The wind speed and wind direction are determined to $v_{\text{w}}=0.42\frac{\text{m}}{\text{s}}$
and $\psi_{\text{w}}=219.92^\circ$, respectively. A vector pointing into that direction is also depicted in the graph of the flight path.

\subsection{Figure-eight patterns}

\begin{figure}[t]
\centering
\includegraphics[width=0.8\textwidth]{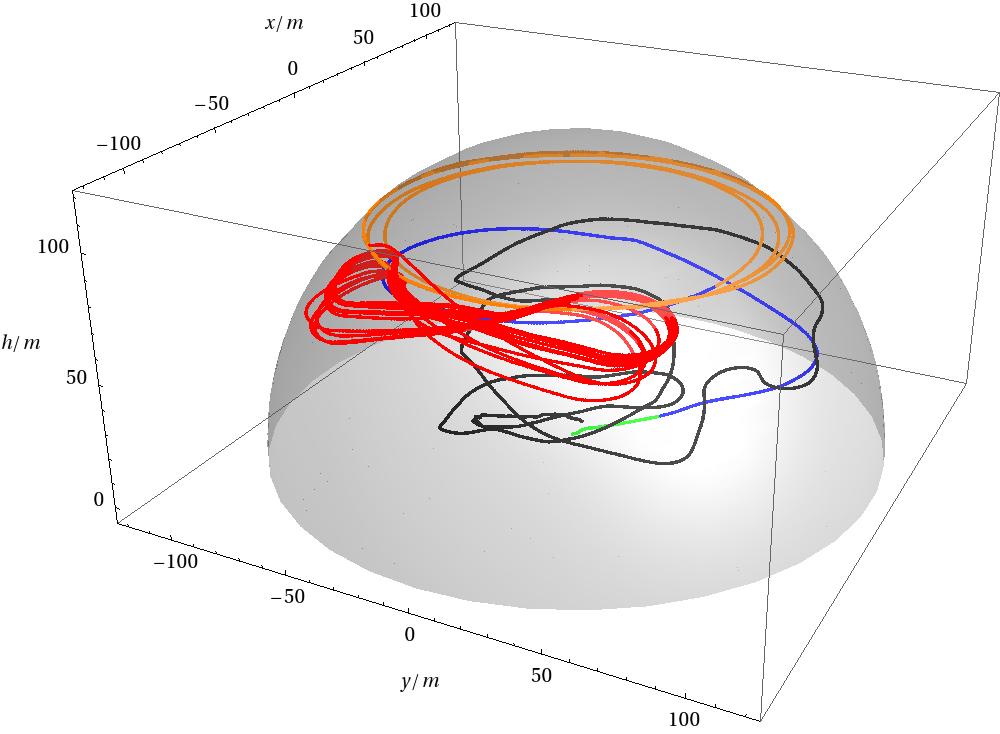}%
\caption{Modes of the tethered test flight. It consists of a launching phase (green),
a helix generated by a horizontal circular movement combined with an  increment
of the altitude (blue), a horizontal circular flight path at the
target altitude (orange), the figure-eight pattern (red) and a landing phase (black). All flight phases apart from the manual landing phase are
autonomous. Entering and leaving the figure eight-pattern is initiated manually.
\label{fig:testflightmodes}
}
\end{figure}

In order to test the figure-eight pattern, the aircraft has to 
be launched, brought to the entry point of the figure-eight pattern and it has also to be landed at the end of the test. Thereby, one has to consider that 
the attached tether constrains the launch and land patterns. The
launch, transfer to the pattern entry point and the pattern itself are
flown in a fully autonomous way. In the current implementation, the
abort from the pattern and the landing procedure are controlled manually.
A test flight consists of 
the mentioned phases, which in figure
\ref{fig:testflightmodes} are highlighted in different colors
for one data set. 
In the following, the quantitative analysis is performed for the figure-eight
flight phases.

\begin{figure}[t]
\centering
\includegraphics[width=0.5\textwidth]{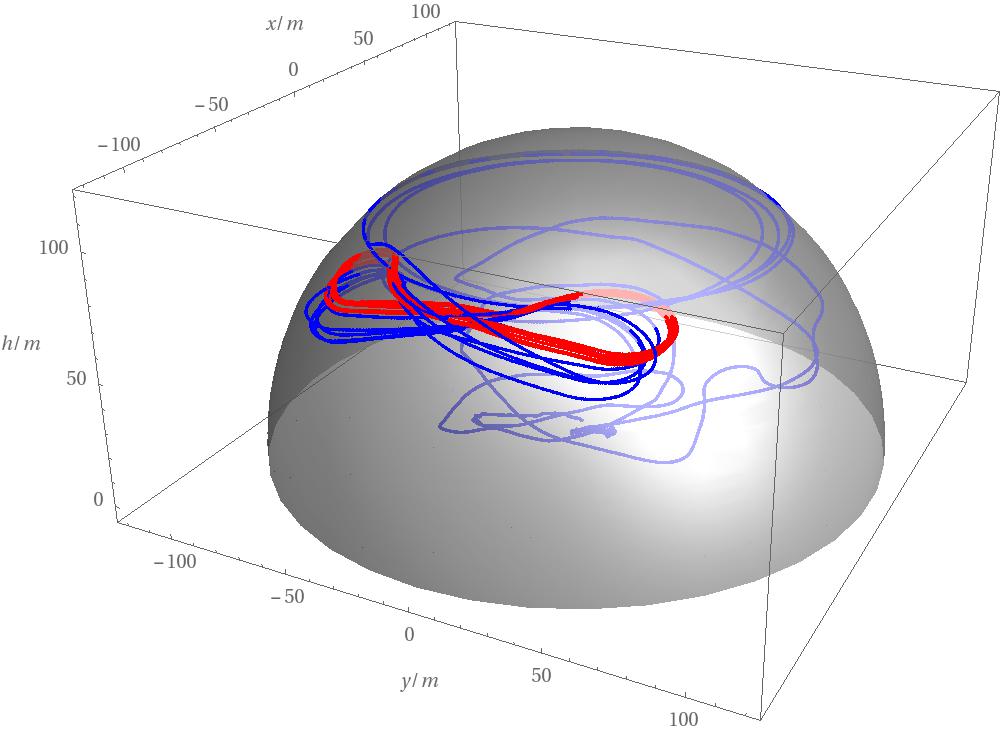}%
\includegraphics[width=0.5\textwidth]{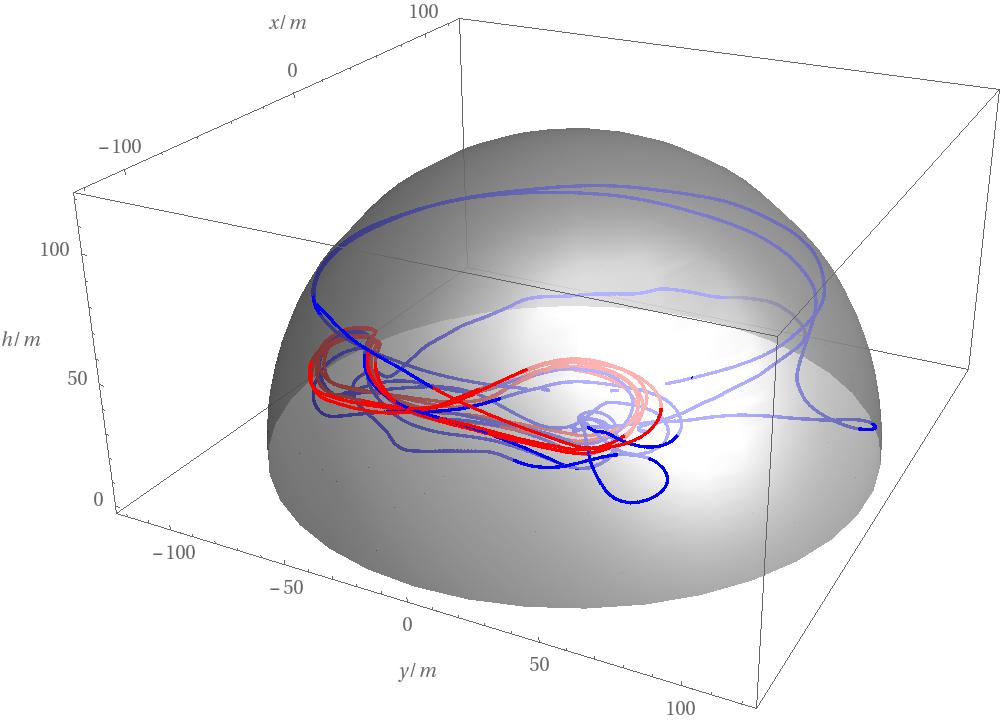}%
\caption{\label{fig:eights}
Flight paths with figure-eight pattern elevated towards south at angles $\gamma=45^\circ$ and $\gamma=30^\circ$. The stationary figure-eight parts that are analysed in detail are highlighted in red.}
\end{figure}

Two flight paths that contain figure-eight patterns 
elevated towards south of the NED coordinate system at elevation 
angles $\gamma=45^\circ$ and $\gamma=30^\circ$ have been recorded.
They are shown in figure \ref{fig:eights} as extracted from the
{\tt NKF1} entries of the corresponding log files.
The shown spheres have radii $R=120\text{m}$, which is the 
distance from the `home' position set in the tethered flight modes.
In the ideal case of no tether drag that would be the length of the tether.
Since the wind 
velocity was only $v_{\text{w}}=0.42\frac{\text{m}}{\text{s}}$, the throttle was permanently activated and the 
aircraft had to be manually landed before it ran out of battery power. 

The periodic stationary flights path segments along
the two figure-eight patterns are analyzed in the following.
They are depicted in red in figure \ref{fig:eights}.
The automatic and manual starting and landing phases, respectively, 
and also the  non-stationary motions in the figure-eight flying mode
are depicted in blue.

\begin{figure}[t]
\centering
\includegraphics[width=0.5\textwidth]{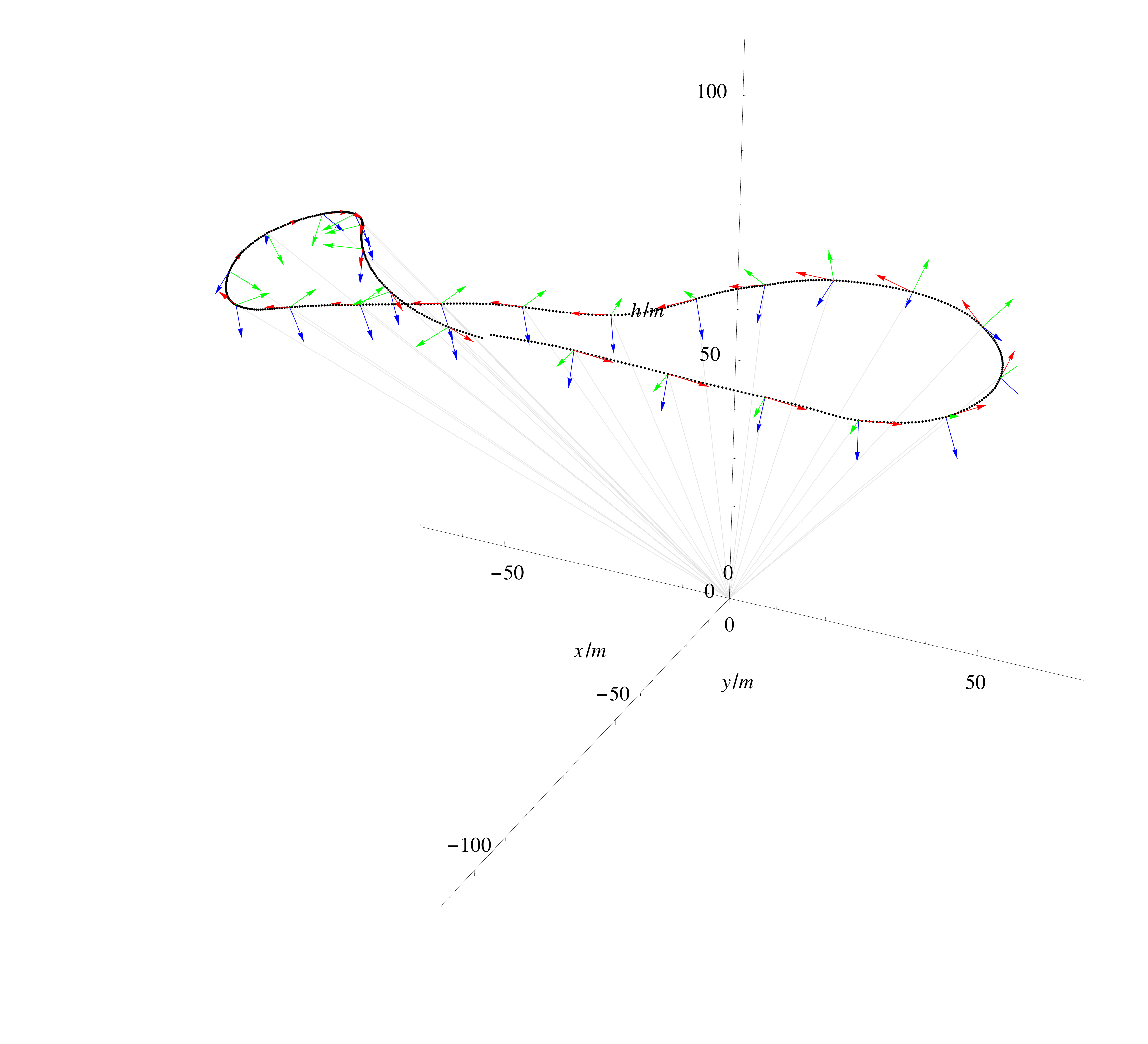}%
\includegraphics[width=0.5\textwidth]{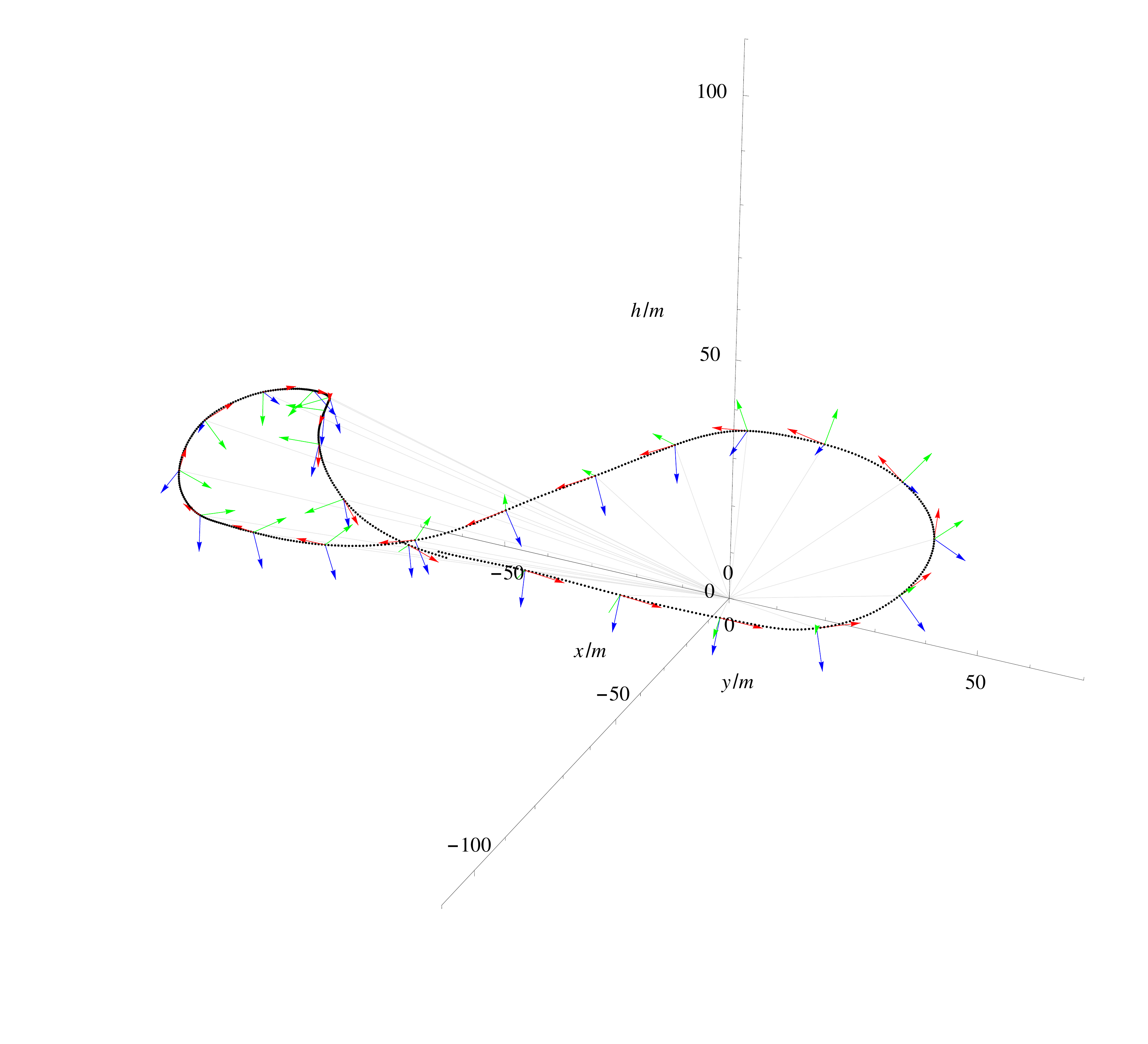}%
\caption{\label{fig:airframe}
Aircraft-fixed coordinate system at some positions of a single figure-eight period at elevation angles $\gamma=45^\circ$ and $\gamma=30^\circ$. The nose, starboard and down axes are depicted in red, green and blue, respectively.
}
\end{figure}

The attitude of the aircraft is also recorded in the {\tt NKF1} entries as 
the three Euler angles roll ($\Phi$), pitch ($\Theta)$ and yaw ($\Psi)$. 
From these angles, the aircraft-fixed coordinate system can be constructed.
In figure \ref{fig:airframe},
it is depicted at some positions of a single figure-eight period.
The eastern and western segments of both figure-eight patterns are different.
A reasonable explanation of this asymmetry is that the figure-eight patterns 
are elevated towards south rather than downwind. 
The downwind location is south-east
as evident from figure \ref{fig:cARSP}.

\begin{figure}[t]
\centering
\includegraphics[width=0.5\textwidth]{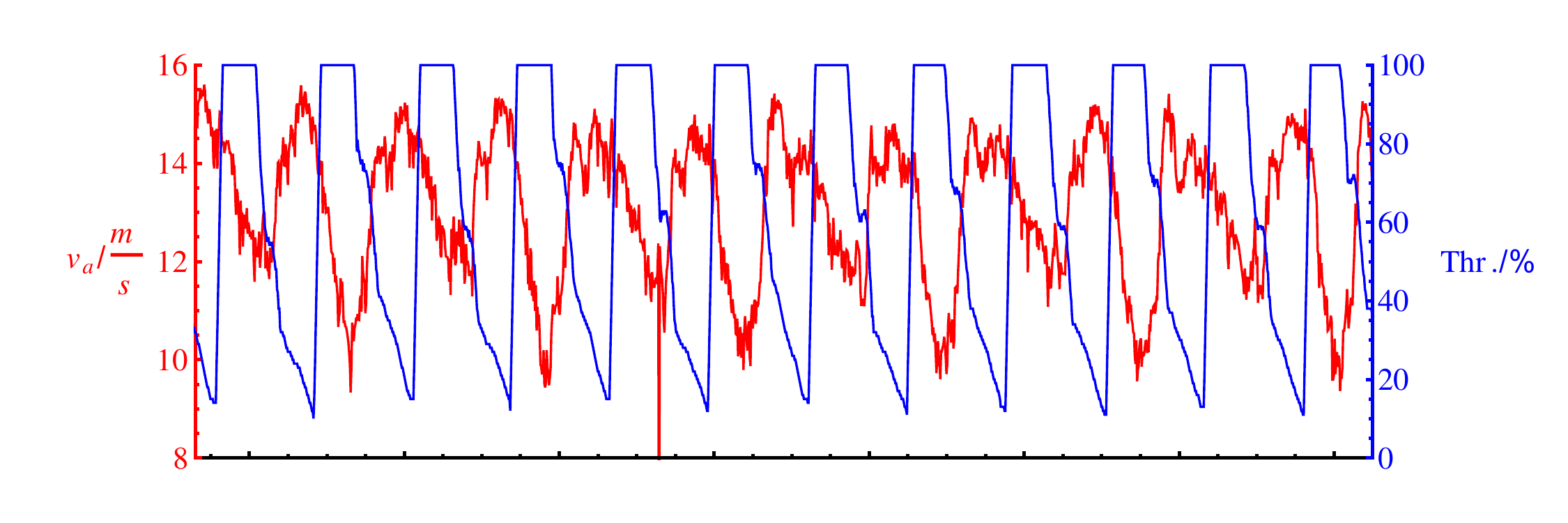}%
\includegraphics[width=0.5\textwidth]{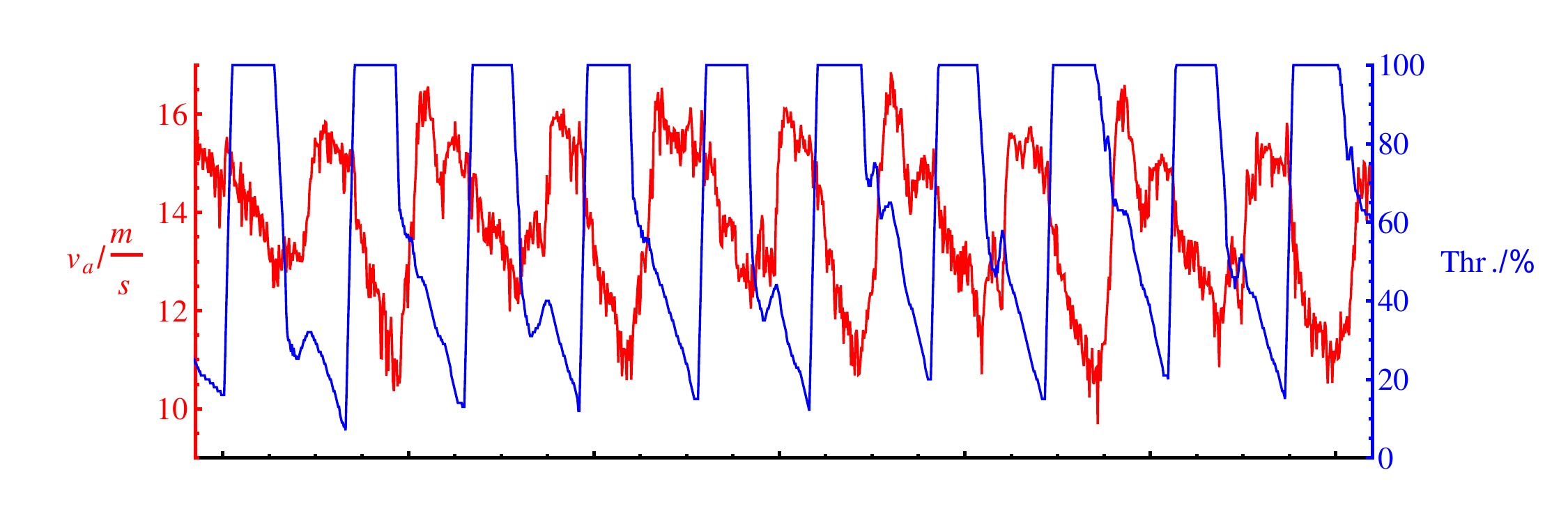}\\
\vspace{-13pt}%
\includegraphics[width=0.5\textwidth]{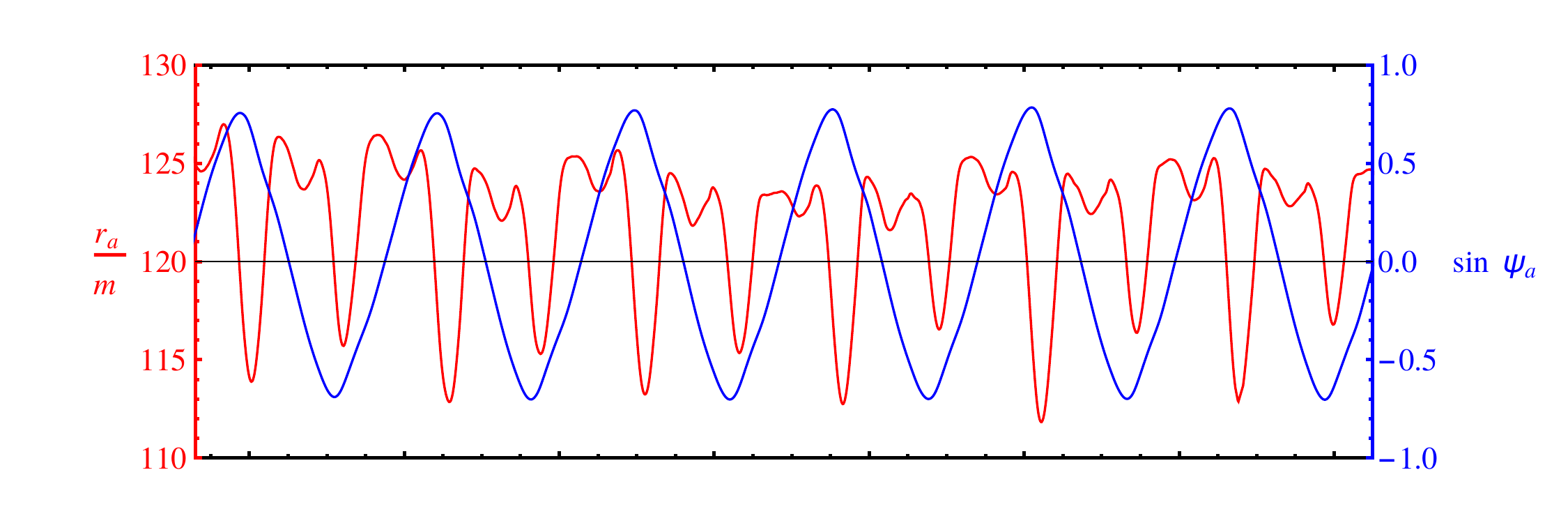}%
\includegraphics[width=0.5\textwidth]{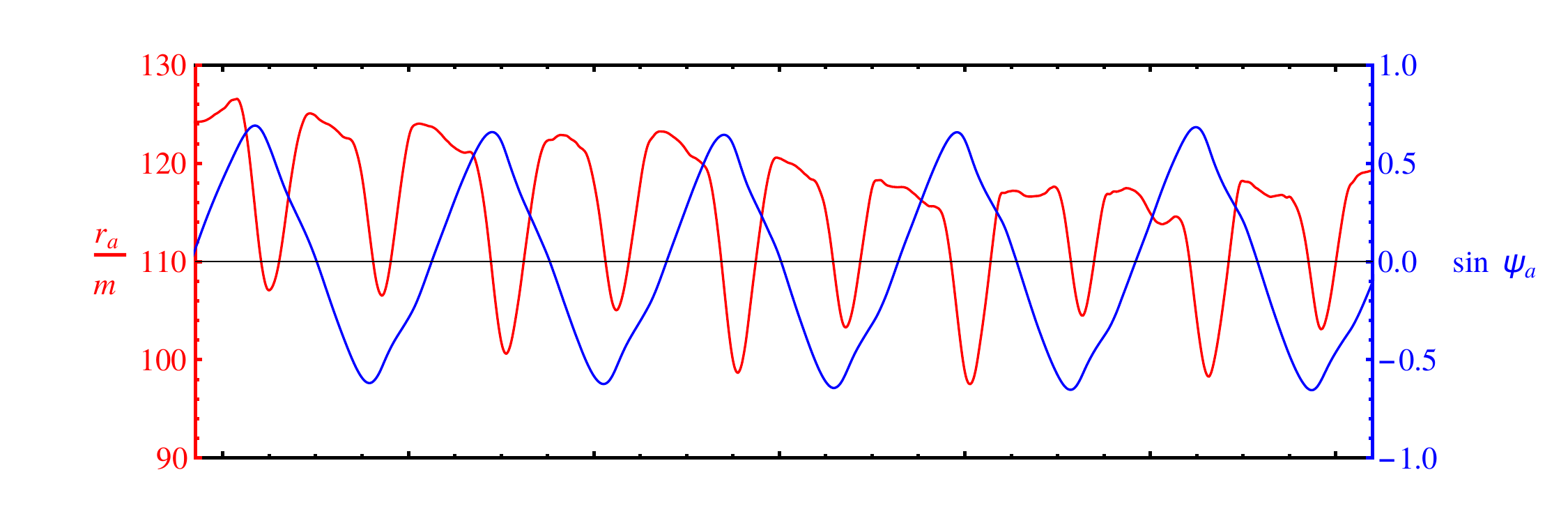}\\
\vspace{-9pt}%
\includegraphics[width=0.5\textwidth]{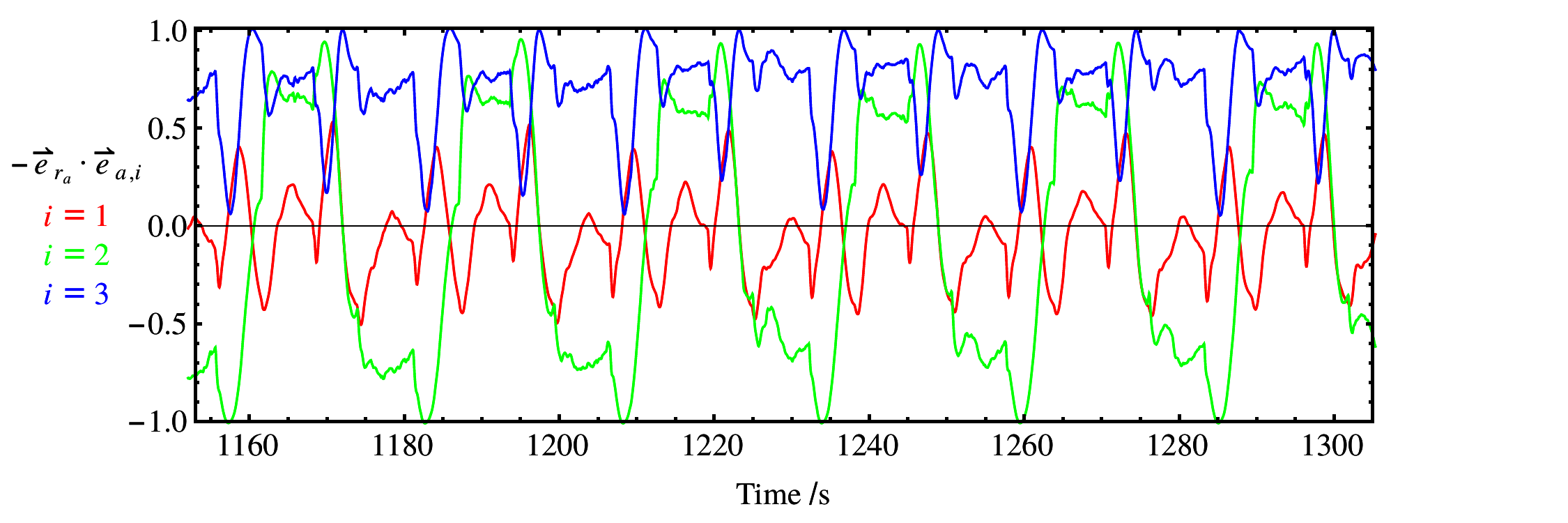}%
\includegraphics[width=0.5\textwidth]{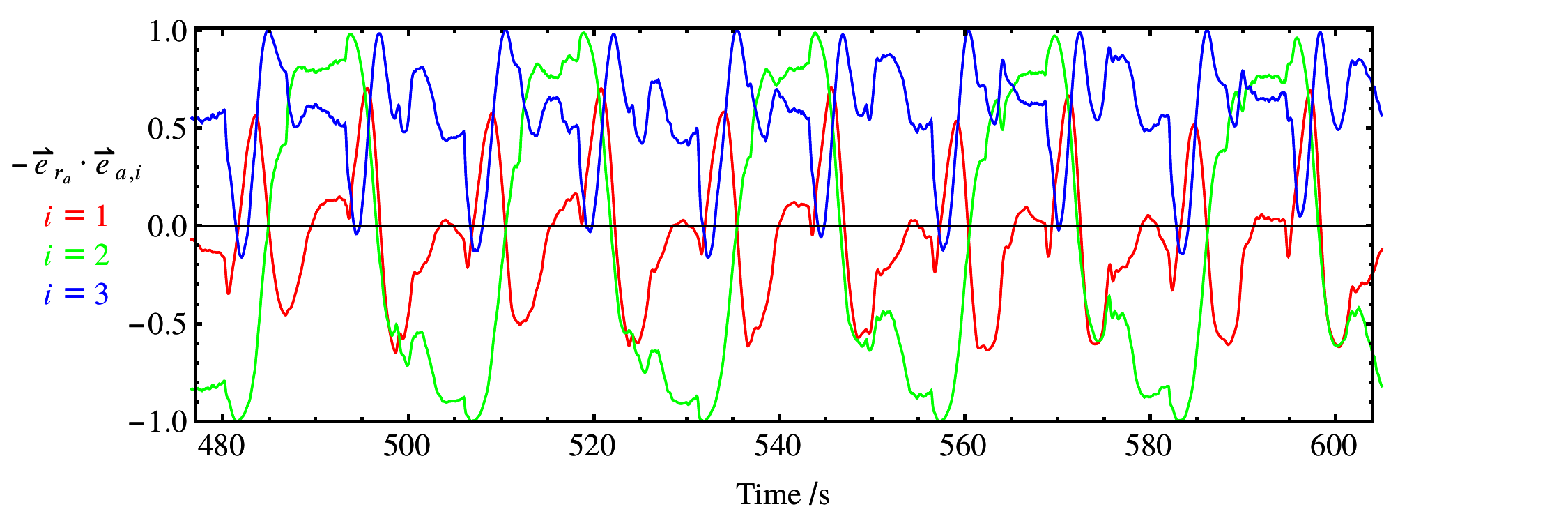}%
\caption{\label{fig:timeanalysisfigeight}
The time dependence of the throttle, airspeed $v_{\text{a}}$, angle $\psi_{\text{a}}$ between the north direction and the lateral projection of the position vector $\vec r_{\text{a}}$ of the aircraft and distance $r_{\text{a}}$. 
Time-dependent projections of the aircraft-fixed frame on the direction
$-\vec e_{r_\text{a}}$ pointing to the `home' location. 
The left and right plots are for the figure-eight patterns at elevation angles $\gamma=45^\circ$ and $\gamma=30^\circ$, respectively.
}
\end{figure}

Since during the two recorded figure-eight flights the wind speed was very low
as also evident from the fit in figure \ref{fig:cARSP}, 
the throttle is automatically activated by the TECS controller.
The time dependence of the throttle and airspeed for both figure-eight patterns
at elevation angles $\gamma=45^\circ$ and $\gamma=30^\circ$ is shown in figure \ref{fig:timeanalysisfigeight}. 
As can be seen, the throttle is activated whenever the airspeed drops.
This periodic behaviour is correlated with the position of the aicraft on the figure-eight pattern. A parameter that captures some of that position information is the angle $\psi_{\text{a}}$\footnote{This angle should not be confused with the angle $\psi_{\text{a}}$ introduced in Section \ref{sec:airspeedcalibration}.} between the lateral projections of the two vectors pointing from the center of the sphere to the crossing point and to the aircraft. At the crossing point
$\sin\psi_{\text{a}}=0$ and at the (eastern and western) apices
$\sin\psi_{\text{a}}$ assumes its maximum and minimum that depends on the extension of the figure-eight pattern. The airspeed 
becomes maximal when the aircraft flies along the geodesic segments, i.e.\ in an interval around the crossing point. The minima
of the airspeed occur approximately at those times where the aircraft reaches one of the two apices, i.e.\ when it is turning upwards in the turning circles.
The airspeed in the eastern and western segment is not symmetric. A reasonable explanation is that the figure-eight pattern is oriented south rather than 
exactly downwind as evident from the reconstructed wind direction shown in figure \ref{fig:cARSP}.

The variation of the airspeed leads to a variation of the lift and drag force.
These forces cannot be measured directly. Instead, the tether force could be measured as highly relevant parameter for power generation. However, no 
force sensor is installed yet. By investigating the variation of the distance
$r_{\text{a}}$ of the aircraft from the `home' location one
can nevertheless show that the tether is under tension and that this
tension depends on the airspeed.
The measured distances are also shown in figure \ref{fig:timeanalysisfigeight}.
One can see that the maxima and minima of the distance follow the 
maxima and minma of the airspeed. The tether, which runs along 
a curve due to tether drag, is hence stretched more if the airspeed and hence
lift force and tether tension increase. Similar to the airspeed also 
the distance is not symmetric between the eastern and western figure-eight segments, possibly caused by the fact that the figure-eight pattern is not 
oriented downwind. 
The asymmetry could be used to automatically adapt the flight path such that 
the figure-eight pattern is always oriented downwind.

Next, we analyze the attitude of the aircraft 
flying along the figure-eight pattern. To this purpose the 
projections of the three aircraft-fixed axes onto the unit vector $-\vec e_{r_{\text{a}}}$ pointing from the position of the aircraft to the `home' position
are shown in figure \ref{fig:timeanalysisfigeight} for both figure-eight patterns. The nose-axis is roughly perpendicular to this vector. The maximal deviations occur when the aircraft passes through the apices of the figure-eight pattern. The down-axis is very roughly pointing to the `home' position. The deviations are maximal whenever the projection of the starboard axis onto 
$-\vec e_{r_{\text{a}}}$ is extremal. This projection depends on the roll angle 
in the tangential frame of the sphere and hence determines the deviation from 
the geodesic flight path. This deviation should ideally vanish but 
is found to be not zero when the aircraft passes through the crossing point of 
the figure-eight pattern.
A possible explanation for this deviation is that the attitude of the aircraft has to be such that the gravitational force 
is compensated. This requires that the aircraft is rotated towards the horizontal attitude more than without the force of gravity.  Moreover, the tether force has not been considered so far when the roll angle is determined according to Newton's law. Especially in strong winds the tether force would dominate 
the gravitational force and hence should reduce the observed deviation.

\begin{figure}[t]
\centering
\includegraphics[width=0.5\textwidth]{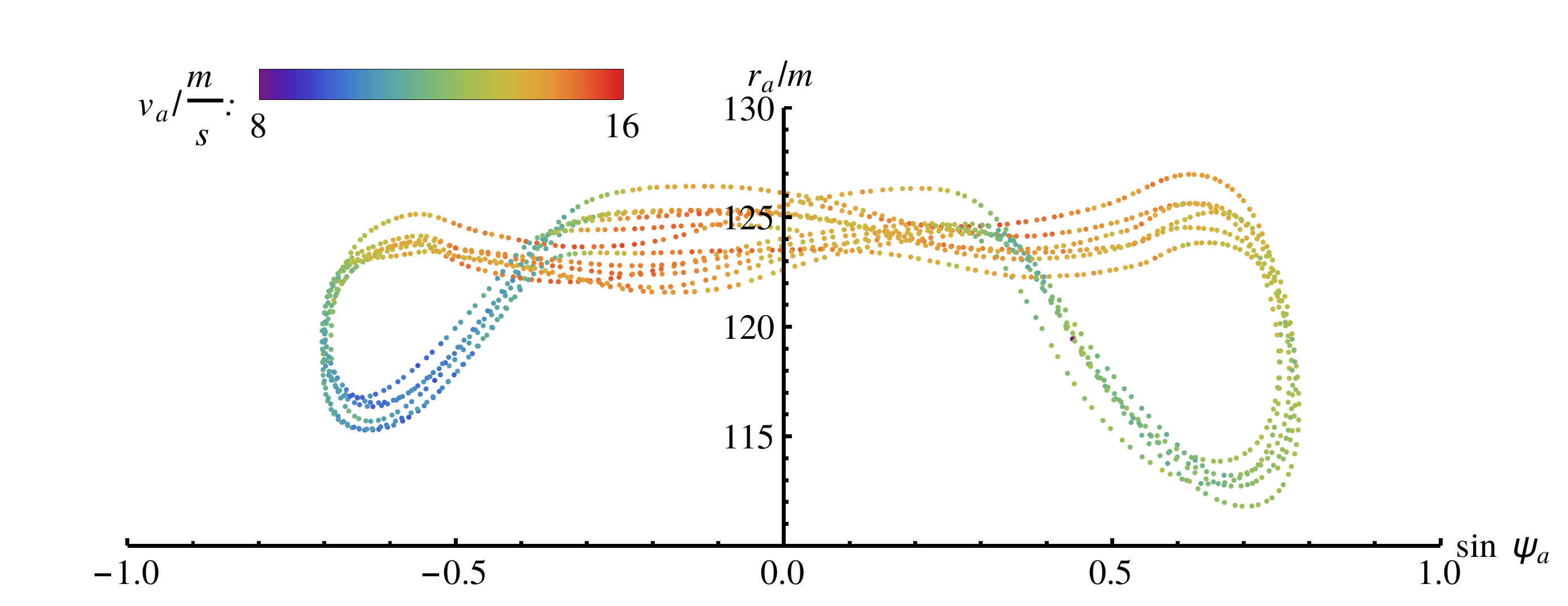}%
\includegraphics[width=0.5\textwidth]{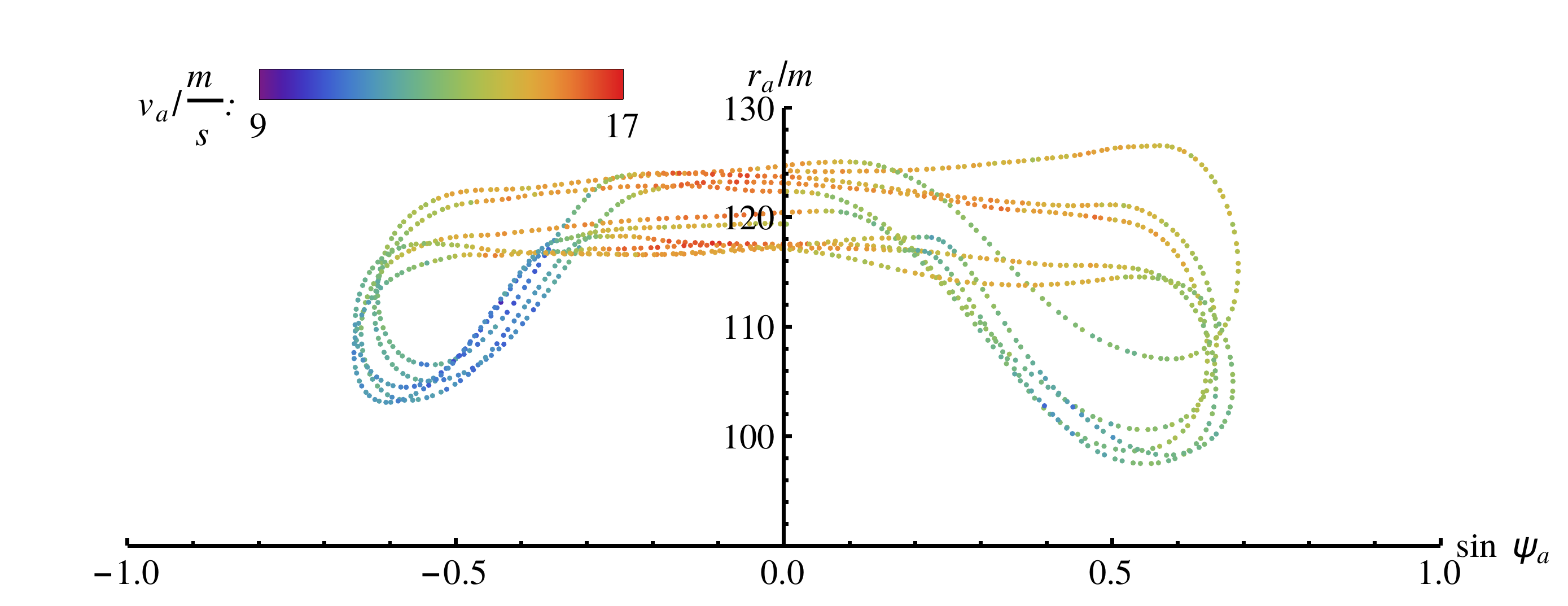}\\
\includegraphics[width=0.5\textwidth]{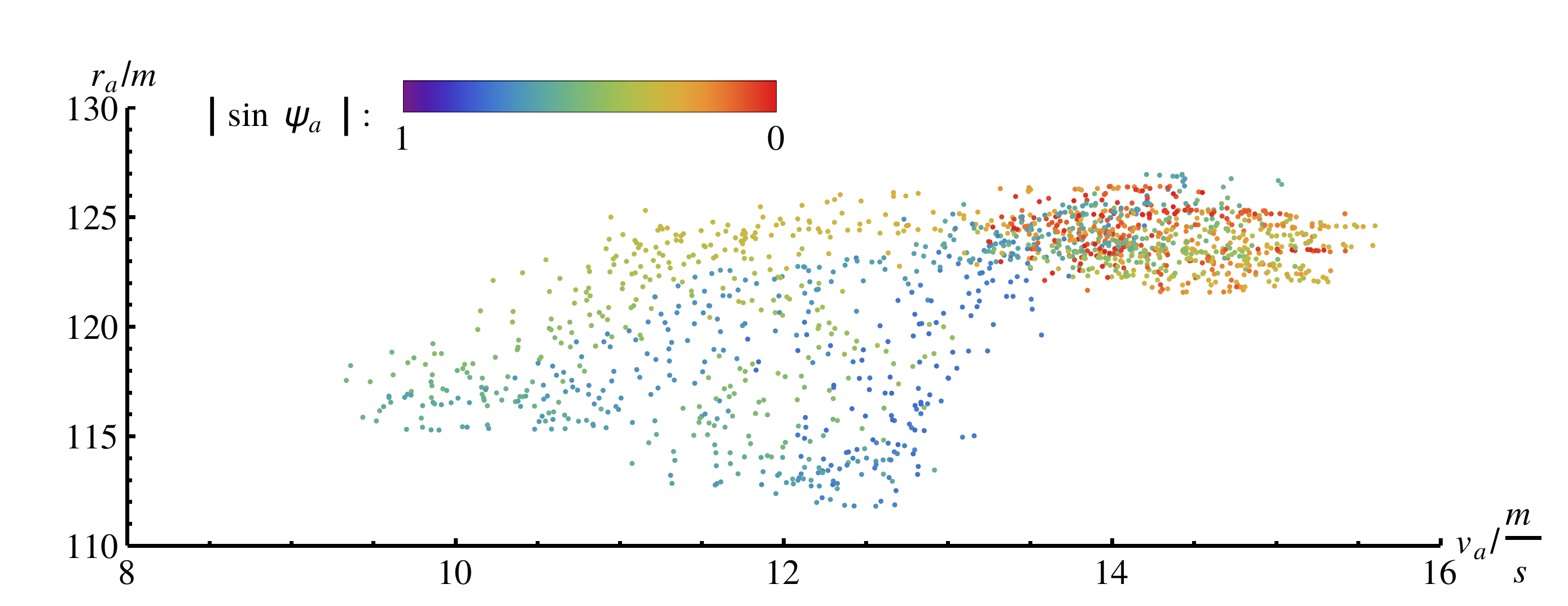}%
\includegraphics[width=0.5\textwidth]{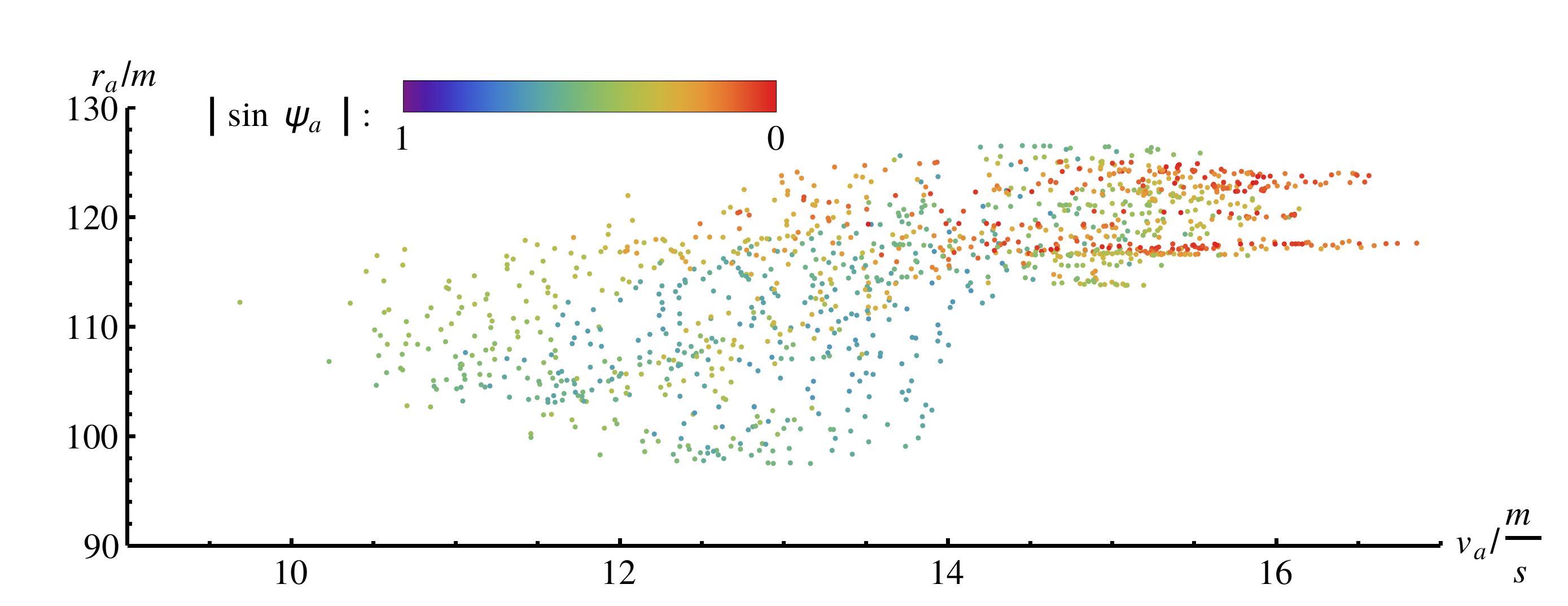}\\
\caption{\label{fig:angleARSPrad}
The distance $r_{\text{a}}$ from the aircraft to the `home' location in dependence of the angle $\psi_{\text{a}}$ between the north direction and the lateral projection of  $\vec r_{\text{a}}$ and the airspeed $v_{\text{a}}$ for the figure-eight patterns
at elevation angles $\gamma=45^\circ$ and $\gamma=30^\circ$.
}
\end{figure}

Finally, it is very interesting to display the distance $r_{\text{a}}$ as 
function of the angle $\psi_{\text{a}}$ and of the airspeed $v_{\text{a}}$. 
The results are shown in figure \ref{fig:angleARSPrad} for both 
figure-eight patterns and with $v_{\text{a}}$ or $|\psi_{\text{a}}|$ 
encoded by the color of the points.
The airspeed  $v_{\text{a}}$ and the distance $r_{\text{a}}$ are maximal on the geodesic segments, and they
are minimal on the turning circles. Moreover, the interval of possible distances that are measured for a given airspeed becomes smaller with increasing airspeed. A reasonable explanation for this is that at higher airspeed and distance the tether is less curved and hence its dynamics has less influence on the 
system. Moreover, the respective upper limits of $r_{\text{a}}$ 
are visible as sharp cutoffs. These limits are related to the 
tether lengths via the unknown tether curve.  
A model of that curve in the stationary case could be used to 
estimate their relation. Moreover, the data for the figure-eight pattern 
elevated at $\gamma=30^\circ$ suggest that two different tether lengths 
were used in this flight, since it seems to consists of two copies of 
a single data set that is shifted along the $r_{\text{a}}$ axis. This is 
indeed the case, during the flight the tether length was slightly changed by 
 a couple of meters in order to optimize the felt tether force and flight pattern.


\section{Conclusions}\label{sec:conclusions}

In this paper we have presented AWEsome, our test platform for airborne wind
energy systems. It consists of low-cost hardware (for which we spent
less than US\${$\,$}1000) and open-source software and hence allows
research groups and small startups to test their design strategies
without large financial investments. We have demonstrated that an
off-the-shelf polystyrene model aircraft with certain reinforcements by
carbon fiber fabric can be used as the airborne part of an AWE system
in cross-wind flight, where the aircraft is connected to the ground
via a tether. We have presented the principal functionality our
implementation of two flight modes where the length of the tether is
fixed and the aircraft hence moves on a hemisphere. As periodic flight
paths we have chosen an inclined circle and a figure-eight pattern.
The latter consists of two great and two small circles and hence is
based on the algorithms of the former.  The SITL environment of the
autopilot software enabled us to debug and test our code prior to
starting field tests.  The results from our real flight tests we have
analyzed in detail. This is possible thanks to a complete logging of 
all sensor and (processed) flight data.
While a force sensor for measuring the tether
tension is not implemented yet, the results show that the aircraft
produces a tether tension that -- if scaled up to a larger plane --
could in principle be used for power production.

Although our test platform is functional, the current implementation
can be extended and improved in many ways. First of all, the
figure-eight pattern that consists of four circle segments is only
$C^1$. The curvature $\kappa$ and geodesic curvature
$\kappa_{\text{g}}$ are discontinuous at the transgression points of
the segments. The roll angle of the aircraft would hence have to be altered
infinitely fast to follow that curve.  In order to avoid this, the
curve should at least be $C^2$ at all points.  For instance, one could
implement generalizations of certain sphere-cylinder intersections
(Viviani's curves) that are $C^\infty$ and encompass figure-eight
patterns.

The separation of the navigation into lateral and speed-height
controller appears not to be well suited if an additional force (the
tether force) is present and the aircraft moves on a hemisphere. So
far, the tether force has not been considered explicitly in the
control algorithms. The incorporation of the tether force into the
algorithms requires a model for the tether curve or sensors that
measure the direction of the tether at the aircraft.  In any case a
sensor that measures the tether force should be added.  The navigation
controller could be split into the directions tangential and normal to
the sphere at the current position of the aircraft. Our expectation is that 
this should be advantageous, at least in case of a strong tether force
that dominates the gravitational force.

Another interesting area of development is the autonomous adaption of
the flight pattern to the wind direction. In \cite{Gehrmann:2016} one
of us has shown that the wind direction can be estimated from the
asymmetry of the flight pattern, which in turn can be used to adapt to
changing winds. Also, the flight pattern itself could be  
adapted in real time in order to maximize the power output.

Last but not least, the ground station that so far consists of an
offshore fishing rod should be replaced by a motor driven drum that
communicates with the ground station and the aircraft in order to
synchronize the release and withdrawal of the tether with the state of
the aircraft.  The ultimate goal is to construct an
automatized launch and landing system.

Concerning further aspects of the test environment, the simulation and
data analysis tools could be improved. This concerns the possibility
to add tether models to the simulation. Moreover, it would be great to
use the flight data in order to extract the aerodynamic coefficients
of the real aircraft, that can then be used to set up a more realistic
model for the simulation.

We plan to tackle these improvements in the near future. In the
meantime, the source code and construction plans of the hard- and software
described in this paper are available upon request. 
We hope that it serves as a basis for close
collaborations with the AWE community.

\section*{Acknowledgements}

First of all we thank Daidalos Capital for financing the hardware of
our test platform.  We thank Thomas Petri who as an experienced model
pilot helped us to perform the first flight tests of our aircraft
without tether, Tobias Schiffer for assistance during several test
flights, and Roland Schmehl, Steffen Schaepe, Peter Wagner and Klaus
Desch for very helpful discussions. We thank Antonello Cherubini for
detailed comments on the manuscript. C.\ S.\ thanks the Experimental
Particle Physics group at Bonn University, where part of this work was
done, for their warm hospitaly.
Our biggest thanks goes to the community of open-source
software development. 
Without the developers' spirit to share their knowledge and 
without the tremendous work they encoded in the free software used 
and adapted here, this project would not have been possible. 

\lohead{\normalfont \headmark}    
\footnotesize

\bibliographystyle{utphys}
\bibliography{ardupilot_ftf_ref.bib}

\end{document}